 \documentclass[final,1p,times]{elsarticle}

\usepackage{amssymb}
\usepackage{amsmath}
\usepackage{mathtools}
\usepackage{bm}

\usepackage[english]{babel}
\makeatletter
\let\oldtheequation\theequation
\renewcommand\tagform@[1]{\maketag@@@{\ignorespaces#1\unskip\@@italiccorr}}
\renewcommand\theequation{(\oldtheequation)}
\makeatother
\usepackage[english,pdfauthor={Marcel Mokbel},  pdfauthor={Sebastian Aland}, pdftitle={Paper}, breaklinks=false, colorlinks=false, linkbordercolor={0 0 1}, linktocpage=true, hyperfootnotes=true]{hyperref}
\usepackage{multirow} 
\usepackage{multicol} 
\usepackage{tabularx} 
\usepackage{longtable} 
\usepackage{array}
\usepackage{todonotes}
\usepackage{float}
\usepackage{graphicx} 
\usepackage{color} 
\graphicspath{{images/}}
\DeclareGraphicsExtensions{.pdf,.png,.jpg} 
\usepackage{subfigure} 
\usepackage[all]{hypcap} 
 \usepackage{tikz}   

  \usetikzlibrary{arrows,calc,snakes}%

\usepackage{paralist}
\usepackage{bibgerm} 
\usepackage{breqn}

\addto\extrasenglish{%
}

\journal{International Journal for Numerical Methods in Fluids}

\begin{document}

\begin{frontmatter}


 \title{An ALE method for simulations of axisymmetric elastic surfaces in flow}
\author[htw]{Marcel Mokbel}
 \author[htw]{Sebastian Aland\corref{cor1}}
 
  \address[htw]{Faculty of Informatics/Mathematics, HTW Dresden, Friedrich-List-Platz 1,  01069 Dresden, Germany}
  \cortext[cor1]{E-mail: sebastian.aland@htw-dresden.de}


\address{}

\begin{abstract}

The dynamics of membranes, shells and capsules in fluid flow has become an active research area in computational physics and computational biology. The small thickness of these elastic materials enables their efficient approximation as a hypersurface which exhibits an elastic response to in-plane stretching and out-of-plane bending, possibly accompanied by a surface tension force.
In this work, we present a novel ALE method to simulate such elastic surfaces immersed in Navier-Stokes fluids. 
The method combines high accuracy with computational efficiency, since the grid is matched to the elastic surface and can therefore be resolved with relatively few grid points.
The focus of this work is on axisymmetric shapes and flow conditions which are present in a wide range of biophysical problems. 
We formulate axisymmetric elastic surface forces and propose a discretization with surface finite-differences coupled to evolving finite elements. We further develop an implicit coupling strategy to reduce time step restrictions.
We show in several numerical test cases that accurate results can be achieved at computational times on the order of minutes on a single core CPU.
We demonstrate two state-of-the-art applications which to our knowledge cannot be simulated with any other numerical method so far: We present first simulations of the observed shape oscillations of novel microswimming shells and the uniaxial compression of the cortex of a biological cell during an AFM experiment. 

\end{abstract}


\end{frontmatter}


\section{Introduction}\label{sec:Introduction}

The dynamics of elastic membranes, shells and capsules in fluid flow has become an active research area in computational physics and computational biology. Example systems include vesicle membranes immersed in fluids \cite{Aland2014b,Baumgart3165,Noguchi2004, Kraus1996}, red and white blood cells transported with the blood plasma \cite{Marth2015, Ye2014, Vlahovska2013,Dupire2012}, general biological cells including cytoplasmic flows \cite{Mokbel, Mietke2018}, or even man-made elastic thin shells to deliver cargo though a fluid \cite{Djellouli2017}. 
Therefore, there exists a broad interest in advanced modeling and simulation technologies that enable the understanding of such systems. 

In all these examples, the elastic material is typically very thin such that its direct numerical resolution with continuum-based discretization techniques requires prohibitively fine mesh sizes \cite{Givoli2004}. 
This problem can be circumvented by replacing volumes of thin layers by dimensionally reduced surfaces, i.e., the elastic material is approximated as a hypersurface of zero thickness.  
Throughout this article, we are particularly interested in elastic surfaces with a finite shear modulus which prevents the surface from strong tangential deformations (while out-of-plane deformation might still be sizable). We therefore explicitly exclude pure lipid vesicles and focus on the more relevant case of cell membranes. The latter are typically connected to a thin elastic actin cortex, the membrane/cortex complex thus forms a thin elastic sheet with strong resistance to surface shear and dilation.
Mechanical properties of the surface include in-plane stretch elasticity and out-of-plane bending elasticity. 
Additionally surface tension forces may arise, for example, in cellular membranes these forces stem from microscopic motor proteins that permanently try to contract the surface \cite{FischerFriedrich2014}. 
Together with the hydrodynamics of the surrounding fluids, all the forces lead to a tightly coupled system of flow and surface evolution. 

Several methods have been developed to numerically simulate such systems. Most popular among them are the boundary integral method, the immersed boundary method and particle collision methods. 
Boundary Element Methods and Boundary Integral Methods couple the Stokes equations to thin shell theory for the elastic surface \cite{Kraus1996,Pozrikidis2003,Veerapaneni2009,Veerapaneni2009b, Boedec2011, Veerapaneni2011, Zhao2011}. The methods are very efficient as they reduce the system to a pure surface problem, yet the limitation to the Stokes regime restricts these methods to small length scales and small flow/shear rates.
An alternative approach are particle methods. Recently a lot of studies used the multiparticle collision dynamics model \cite{Noguchi2007, Paulose2012, Winkler2014} to describe elastic capsules. This numerical scheme is very flexible but remains partly phenomenological.
The immersed boundary method in its original form couples different numerical grids for the surface and the fluid domain by use of smeared-out delta functions
\cite{Le2009, HuJCP2014, Givelberg2004}.
Main problems of such methods are the loss in accuracy associated to this interpolation between grid and the handling of high viscosity ratios \cite{Fai2013,Peskin2002}. These problems can be overcome by coupling immersed boundaries with technically far more complicated cut cell methods \cite{Han2019}.

Alternatively to the above methods, interface capturing methods can be used to track interface movement. The most prominent of which are the level set method \cite{Maitre2009, Salac2011, Laadhari2014, Maitre2012, Doyeux2012} and the phase field method \cite{aland2012benchmark, Aland2014b, Marth2015, lowengrub, Aland2017}.
However, the inclusion of shear and dilational surface elasticity is traditionally not considered in these approaches, as it is not clear how to carry the reference coordinates along the elastic structure. Notably, some first steps have been done in this direction recently for level set \cite{Cottet2006} and phase-field methods \cite{abels2018}. 

The Arbitrary Lagrangian-Eulerian (ALE) method is another method to discretize moving domains and moving boundaries. The method uses a body-fitted grid to couple the advantages of Eulerian and Lagrangian description of the material. The main drawback of these methods is the necessity of re-triangulation when strong grid deformations are involved. 
While the ALE method is the standard method for the interaction of 3D elastic structures with surrounding fluids \cite{Souli2000}, we are not aware of an ALE method for elastic surfaces immersed in fluids. 
The reason for the dominance of ALE methods in 3D fluid-structure interaction (FSI) problems is their high accuracy representation of the domains by a body-fitted grid, their efficiency and the simplicity of implementation. 
In this paper, we aim to propose such a method for elastic surfaces which shares these advantages. We restrict our model to the axisymmetric setting which applies to many surfaces in nature and technology and reduces the system effectively to a two-dimensional problem. This makes the method particularly attractive for problems involving long-term computations (many time steps) or extremely fine grids, for which full 3D simulations are illusive.
Along these lines an axisymmetric Boundary Element Method \cite{trozzo2015axisymmetric} and an axisymmetric Immersed Boundary Method \cite{HuJCP2014} have been proposed recently to simulate vesicles and elastic hyper-surfaces. 
Here, we propose an axisymmetric ALE method, which complements these methods by some distinct advantages.

Axisymmetric ALE methods have been developed and benchmarked in the context of two-phase flows, see e.g. \cite{Li2005, Aland2013}, proving superior efficiency and accuracy in comparison to other two-phase flow methods. Our work adds the elastic forces at the material interface to obtain an efficient and accurate method for elastic surfaces in fluid flows. 
Accordingly we provide a formulation for the elastic surface forces and propose a discretization with surface finite-differences which is coupled to evolving finite elements of the bulk problems. We further develop an implicit coupling strategy to reduce time step restrictions.
While the method is presented here for surfaces which are surrounded by fluids from both sides, it is straightforward to switch off one of the fluid phases, which offers a further advantage w.r.t. Immersed Boundary Methods. 
Similarly, the method can easily account for completely different physics in both surrounding domains. 
We illustrate this by simulating the uniaxial compression of a biological cell filled with intracellular fluid and the propulsion of novel microswimmers having compressible/incompressible fluids inside/outside their elastic shell. 

The rest of this article is structured as follows. In \autoref{sec:axi} the governing equations for the model are introduced along with the suitable boundary conditions to ensure force balance across the membrane. Details about the membrane forces are provided in \autoref{sec:shellMembraneForces}. The numerical discretization of the problem is presented in \autoref{NumericalScheme}. Numerical test cases are presented in \autoref{sec:NumericalTests}. Some details on the implementation can be found in the appendix.

\section{Governing equations} \label{sec:axi}
\begin{figure}
	\centering
	\subfigure[]{\includegraphics[width=0.49\textwidth]{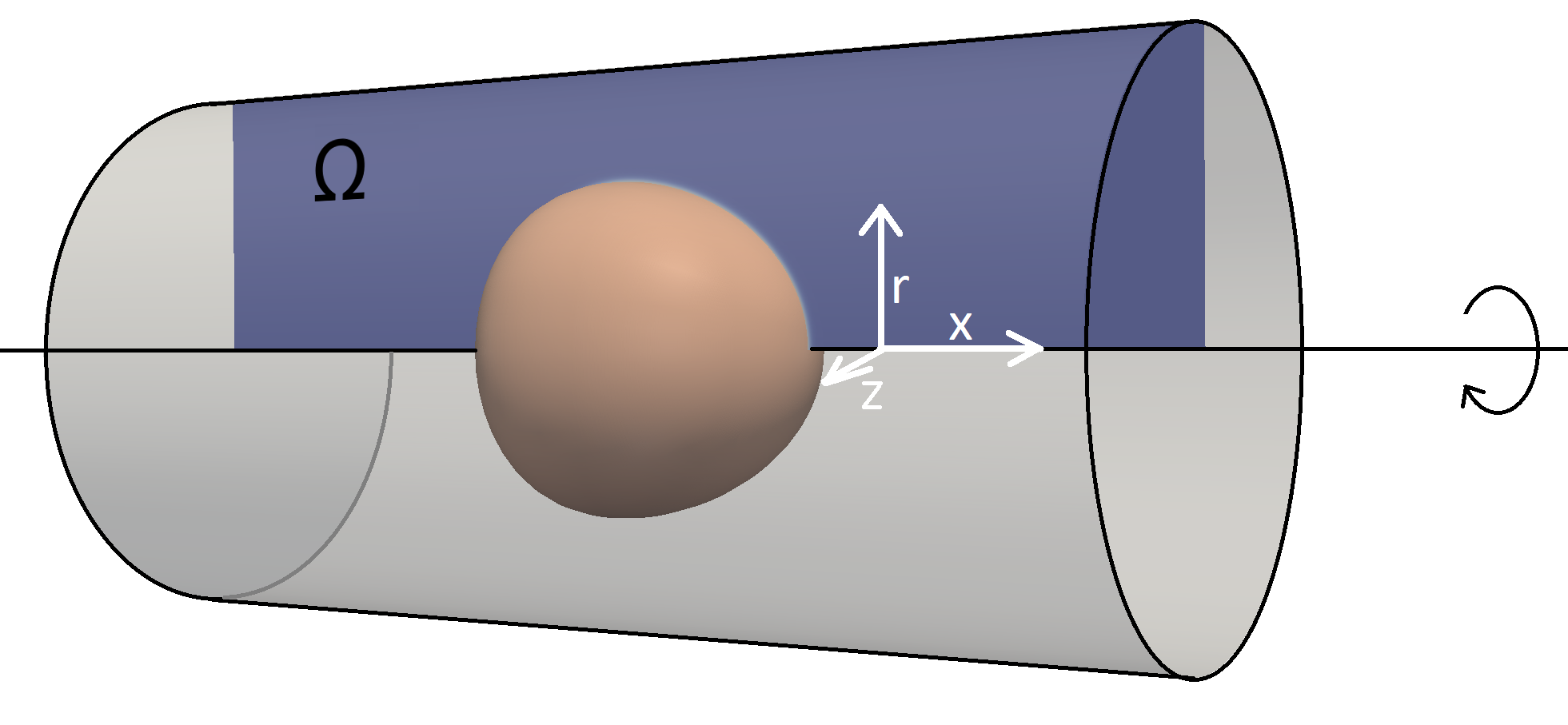}\label{fig:axisymmetric-a}}\hspace{0.25cm}
	\subfigure[]{\includegraphics[width=0.45\textwidth]{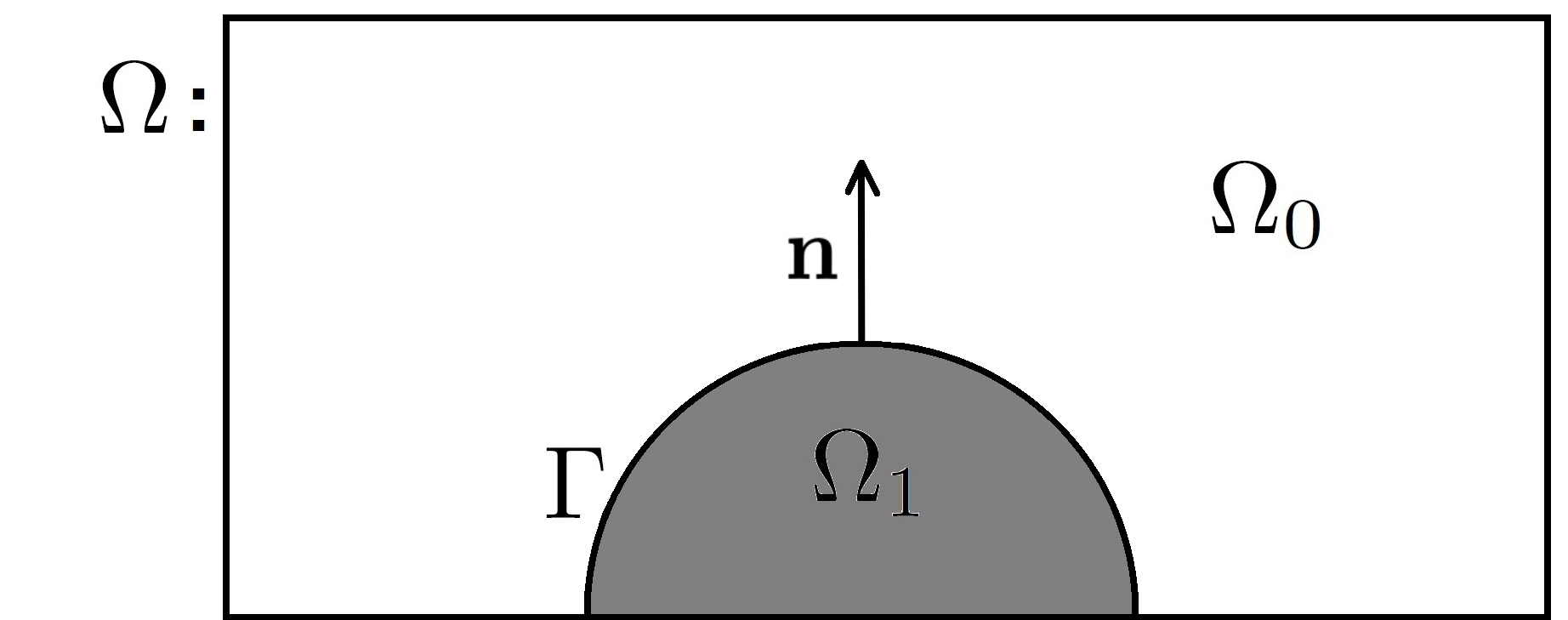}\label{fig:axisymmetric-b}}
	\caption[]{(a) A 3D cylindrical domain containing the elastic shell is considered. The elastic shell is assumed to be a surface of revolution. The domain $\Omega$ is half of the cross section of the cylindrical domain. (b) The axisymmetric domain is composed of the domain for the external fluid $\Omega_0$ and the internal fluid $\Omega_1$ and the elastic membrane $\Gamma$.}
\end{figure}

Consider a (closed) surface of revolution immersed in a cylindrical 3D domain, representing an elastic membrane immersed in a fluid  (\autoref{fig:axisymmetric-a}). Half of the domains cross section acts as the two-dimensional computational domain $\Omega$ (\autoref{fig:axisymmetric-b}). $\Omega$ is composed of two separate parts, the exterior $\Omega_0(t)$ and the interior $\Omega_1(t)$, both describing incompressible viscous fluids: $\Omega={\rm int}(\overline{\Omega_0 \cup \Omega_1})$.
The interface $\Gamma(t)=\overline{\Omega_0(t)}\cap\overline{\Omega_1(t)}$, which separates both domain parts, corresponds physically to an elastic shell and is described by the parametric form
\begin{align}
\mathbf{X}(s,t) = \left(X(s,t),R(s,t)\right)^T\, ,
\end{align}
with the arc length parameter $s$. The $x-$axis is chosen to be the symmetry axis and the $r-$axis is the distance to the symmetry axis in the considered 2D meridian $x-r$ domain. With the divergence operator for cylindrical coordinate systems of the form \cite{abels2018}
\begin{align}
\tilde{\nabla}\cdot = \left(\partial_{x},\frac{1}{r}+\partial_{r}\right),
\end{align}
the Navier-Stokes equations for the external ($i=0$) and internal ($i=1$) fluid read
\begin{align}
\rho_i\left(\partial_t\mathbf{v} + \mathbf{v}\cdot\nabla\mathbf{v}\right) +\nabla p_i &= \tilde{\nabla}\cdot\left( \eta_i\left(\nabla\mathbf{v} + \left(\nabla\mathbf{v}\right)^T\right) \right) + \frac{2\eta_i}{r^2}v_r\cdot  \begin{pmatrix} 0 \\ 1 \end{pmatrix}, && \text{in}\ \Omega _i\label{eq:NavierStokes}\\
\tilde{\nabla}\cdot\mathbf{v} &= 0, && \text{in}\ \Omega_i \label{eq:incompressibility}
\end{align}
with the velocity $\mathbf{v}=\left(v_x,v_r\right)^T$ in $\Omega\cup\Gamma$, pressures $p_i$, viscosities $\eta_i$, and densities $\rho_i$ in $\Omega_i$, respectively. The velocity is continuous across $\Gamma$ and can hence be defined in the whole domain, while the pressure is discontinuous across $\Gamma$ and has therefore to be defined separately in both phases. 
The following jump condition holds at the interface to ensure the force balance at the elastic membrane
\begin{align}
\left[-p_i\mathbf{I} + \eta_i\left(\nabla\mathbf{v} + \left(\nabla\mathbf{v}\right)^T\right)\right]_{\Gamma}\cdot\mathbf{n} &=  -\frac{\partial E}{\partial\Gamma} ,\label{eq:boundaryConditions}
\end{align}
where $\mathbf{n}$ is the interface normal pointing to $\Omega_0$ and $\left[f\right]=f_0-f_1$ denotes the jump operator across the interface $\Gamma$.
The surface force term $\frac{\partial E}{\partial\Gamma}$ is defined in the following. 
At the symmetry axis we specify the usual free slip condition. 

\section{Shell membrane forces}\label{sec:shellMembraneForces}
The membrane is assumed to be an isotropic thin shell with a shell thickness $d$. The force response of a membrane to elastic deformations can be described with in-plane and out-of-plane forces acting to minimize the corresponding elastic energies. The in-plane energy is also referred as stretching energy ($E_{\text{stretch}}$) and the out-of-plane energy as bending energy ($E_{\text{bend}}$). Additionally, a surface tension energy ($E_{\text{tension}}$) can be present. The membrane forces are then given by the first variation of these energies with respect to changes in $\Gamma$
\begin{align}
\frac{\partial E}{\partial\Gamma} = \frac{\partial E_{\text{tension}}}{\partial\Gamma}+ \frac{\partial E_{\text{bend}}}{\partial\Gamma} +\frac{\partial E_{\text{stretch}}}{\partial\Gamma}\, . \label{eq:dEdGamma}
\end{align} 
The surface tension energy, tending to  minimize the membrane area, reads \cite{Mokbel}
\begin{align}
E_{\text{tension}} = \int_{\Gamma}\gamma\, \text{d}A\, ,
\end{align}
with the material specific surface tension $\gamma\ [N/m]$, which is assumed to be constant on the whole membrane.
 
Bending stiffness is the resistance of the membrane against bending (out-of-plane) deformations. The bending energy tends to minimize the deviation of the local curvature from the material's reference curvature (also termed spontaneous curvature). The bending energy reads 
\begin{align}
E_{\text{bend}} = \int_{\Gamma}\frac{c_b}{8}\left(\kappa-\kappa_{\text{ref}}\right)^2\, \text{d}A\, ,
\end{align}
with the material specific bending stiffness $c_b\ [Nm]$ and the total curvature $\kappa = \nabla\cdot\mathbf{n}$, which is twice the mean curvature of the membrane. The spontaneous curvature $\kappa_{\text{ref}}$ is in many practical cases either zero or the total curvature in the initial state, depending on the physical context of the problem.

\begin{figure}
\centering
\subfigure[]{\includegraphics[height=0.22\textwidth]{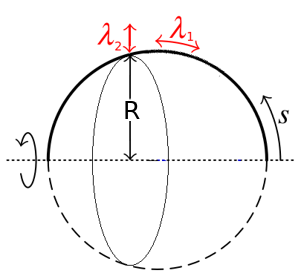}\label{fig:prinicpalStretches}}\hspace{0.25cm}
\subfigure[]{\includegraphics[height=0.18\textwidth]{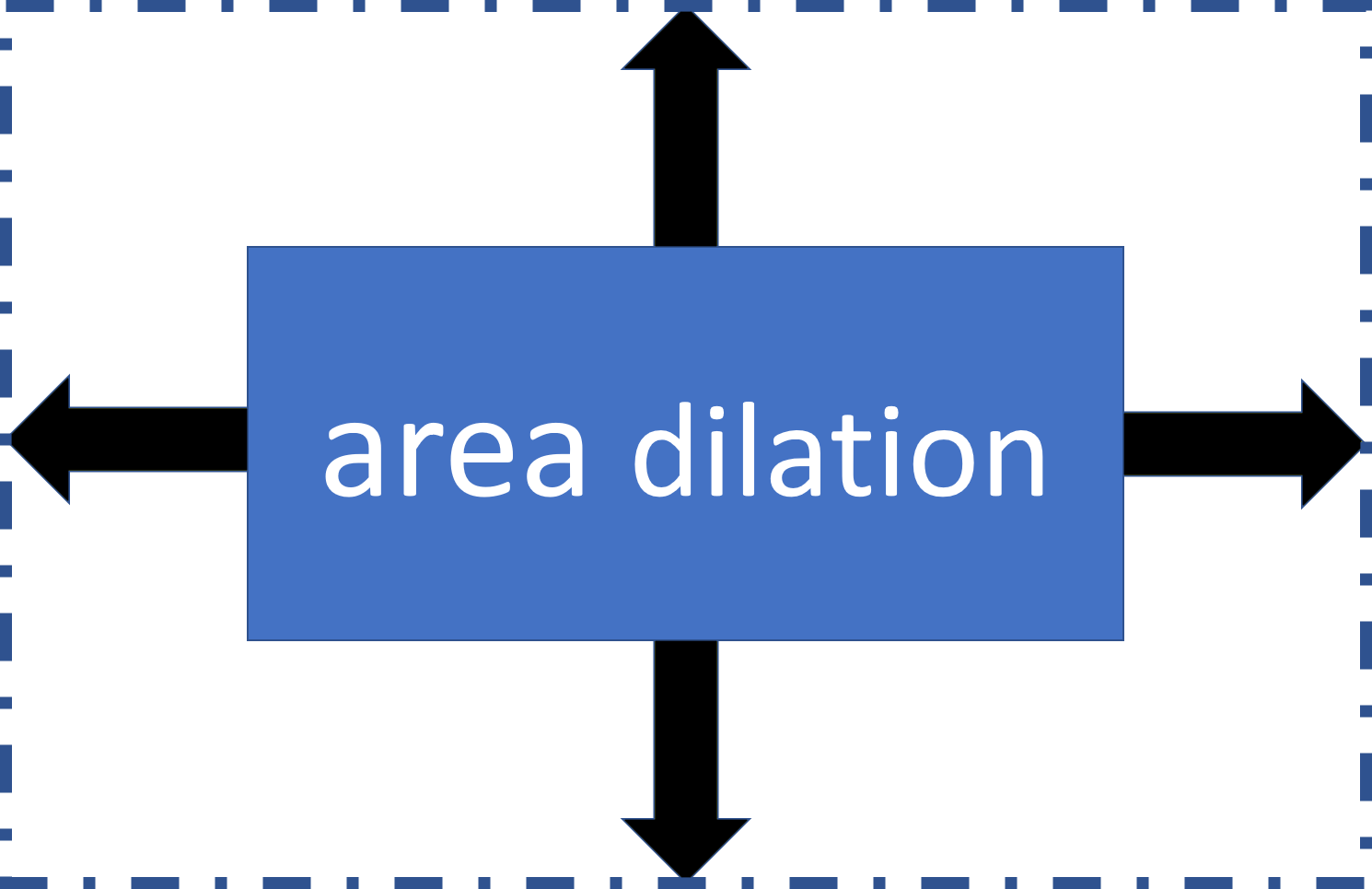}\label{fig:dilation}}
\hspace{0.25cm}
\subfigure[]{\includegraphics[height=0.18\textwidth]{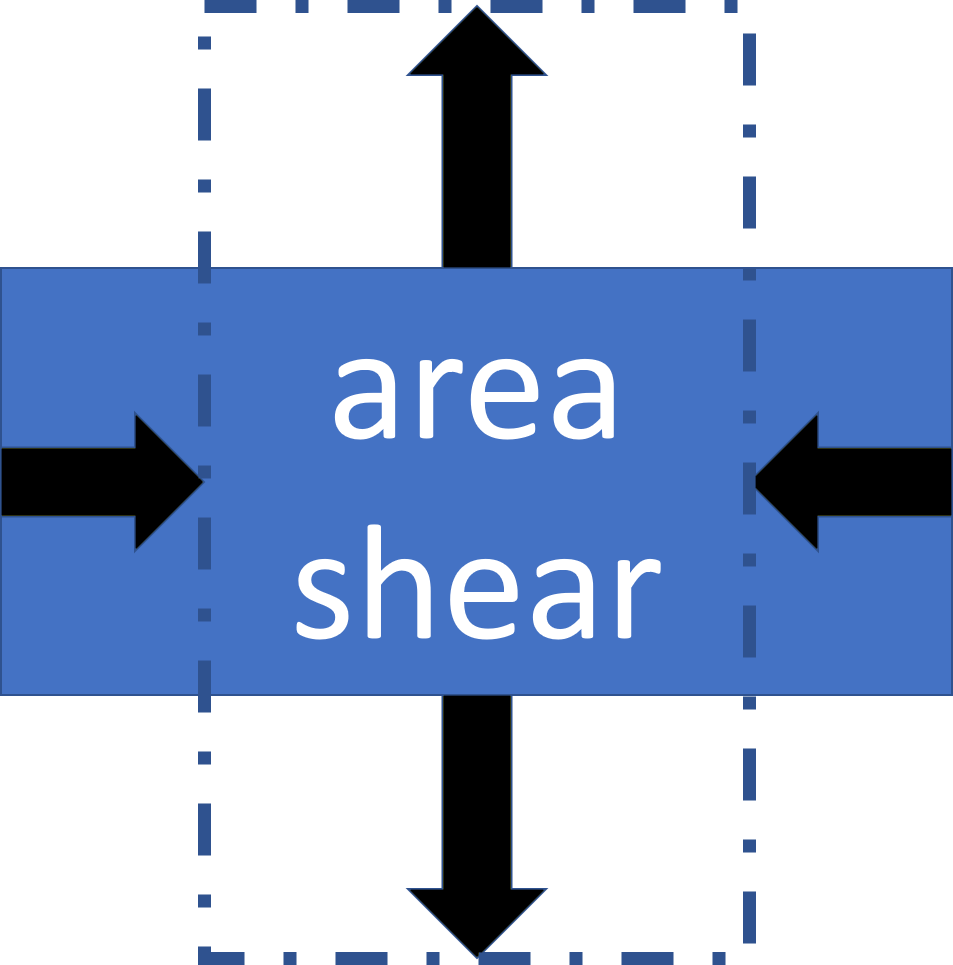}\label{fig:shear}}
\caption[]{(a) Illustration of the principal stretches $\lambda_1$ and $\lambda_2$. (b) Illustration of area dilation. The surface element (blue) is stretched with constant aspect ratio and increasing area. (c) Illustration of area shear. The surface element (blue) is stretched with constant area and non constant aspect ratio. }
\end{figure}

The stretching energy $E_{\text{stretch}}$ minimizes in-plane stretching and compression of the membrane compared to the reference state. In the axisymmetric setting, it is useful to describe the stretching energy in terms of the two principal stretches $\lambda_1$ and $\lambda_2$, which provide information about relative changes of surface lengths in lateral and circumferential direction, respectively. An illustration can be found in \autoref{fig:prinicpalStretches}. The principal stretches read
\begin{align}
\lambda_1=\frac{\text{d}s}{\text{d}s_{\text{ref}}}, \quad \lambda_2=\frac{R}{R_{\text{ref}}},
\end{align}
with the subscript $_{\text{ref}}$ corresponding to the quantities at the same material point in the reference state. 

In 3D elasticity theory, the response of an isotropic elastic body to elastic deformations can be described by two material specific parameters: Young's modulus $E$ and the Poisson ratio $\nu$. 
For a thin elastic material of thickness $d$, these parameters are typically reformulated into surface parameters, for example the the area dilation modulus $K_A$ and area shear modulus $K_S$ \cite{poisson2019}. Considering a rectangular surface element, $K_A$ describes the response of the membrane to in-plane area changes with constant aspect ratio of the surface element (\autoref{fig:dilation}). $K_S$ provides information about the response to in-plane shear deformations with constant area of the surface element (\autoref{fig:shear}). The two parameters together with $c_b$ can be calculated directly from Young's modulus, Poisson ratio and shell thickness:
\begin{align}
K_A = \frac{dE}{2(1-\nu)},\quad
K_S = \frac{dE}{2(1+\nu)},\quad
c_b = \frac{d^3E}{24(1-\nu^2)}
\label{eq:moduli}
\end{align}
In terms of $K_A$ and $K_S$, the stretching energy can be written as  \cite{Mokbel}
\begin{align}
E_{\text{stretch}} = \int_\Gamma \frac{K_A+K_S}{2}\left(\left(\lambda_1-1\right)^2+\left(\lambda_2-1\right)^2\right) + \left(K_A-K_S\right)\left(\lambda_1-1\right)\left(\lambda_2-1\right) ~\text{d}A\, .
\end{align}
The corresponding forces to the respective energies can be obtained by calculating the first variations of the three surface energies:
\begin{align}
\frac{\partial E_{\text{tension}}}{\partial\Gamma} &= \gamma\kappa\,\mathbf{n}\, ,\label{eq:surfaceTensionForce} \\
\frac{\partial E_{\text{bend}}}{\partial\Gamma} &= c_b\left[\Delta_{\Gamma}(\kappa - \kappa_{\text{ref}}) + (\kappa^2 - 2K)(\kappa-\kappa_{\text{ref}}) - \frac{1}{2}\kappa(\kappa-\kappa_{\text{ref}})^2\right]\cdot\textbf{n}\, ,\label{eq:bendingForce}\\
\frac{\partial E_{\text{stretch}}}{\partial\Gamma} &= \left(\kappa\,\textbf{n} -\nabla_{\Gamma}\right)\left[\left(K_A+K_S\right)\left(\lambda_1 - 1\right) + \left(K_A-K_S\right)\left(\lambda_2 - 1\right)\right] - \frac{2K_S}{R}\left(\lambda_1-\lambda_2\right)\begin{pmatrix} 0 \\ 1 \end{pmatrix}\, , \label{eq:stretchingForce}
\end{align}
where $K$ is the Gaussian curvature, $\Delta_{\Gamma}$ the surface Laplacian, and $\nabla_{\Gamma}$ the surface gradient, respectively. Derivations can be found in \cite{Mokbel,lowengrub}.

\section{Numerical scheme}\label{NumericalScheme}

\subsection{Time discretization}
The problem is discretized in time with equidistant time steps of size $\tau$. 
The complete system is split into several subproblems which are solved subsequently in each time step. 
At first the membrane forces are computed from the current membrane shape. Afterwards, the Navier-Stokes equations are solved by an implicit Euler method using the previously computed membrane forces. The resulting velocity is then used to advect the surface and the domains. 

Using the ALE approach, the material derivative of the velocity $\partial^{\bullet}_t\mathbf{v} = d_t\mathbf{v} + \mathbf{v}\cdot\nabla\mathbf{v}$ is discretized in the $n$-th time step as follows
\begin{align*}
\partial^{\bullet}_t\mathbf{v}^n = \frac{\mathbf{v}^n-{\mathbf{v}}^{n-1}_{\text{moved}}}{\tau} + \left(\mathbf{v}^{n-1}-\mathbf{v}^{n-1}_{\text{grid}}\right)\cdot\nabla\mathbf{v}^n,
\end{align*}
where $\mathbf{v}^{n-1}_{\text{grid}}$ is the velocity of the grid movement from the previous time step calculated with one of the mesh smoothing algorithms presented in \autoref{sec:meshMovement}. This term has to be subtracted from the convection term due to the mesh update. The velocity ${\mathbf{v}}^{n-1}_{\text{moved}}$ is the velocity of the last time step, but after the mesh update, i.e. the grid point coordinates have been moved without changing the velocity values in each degree of freedom (DOF).

\subsection{Space Discretization}\label{sec:meshSmoothing}
\begin{figure}
\centering
\includegraphics[width=0.8\textwidth]{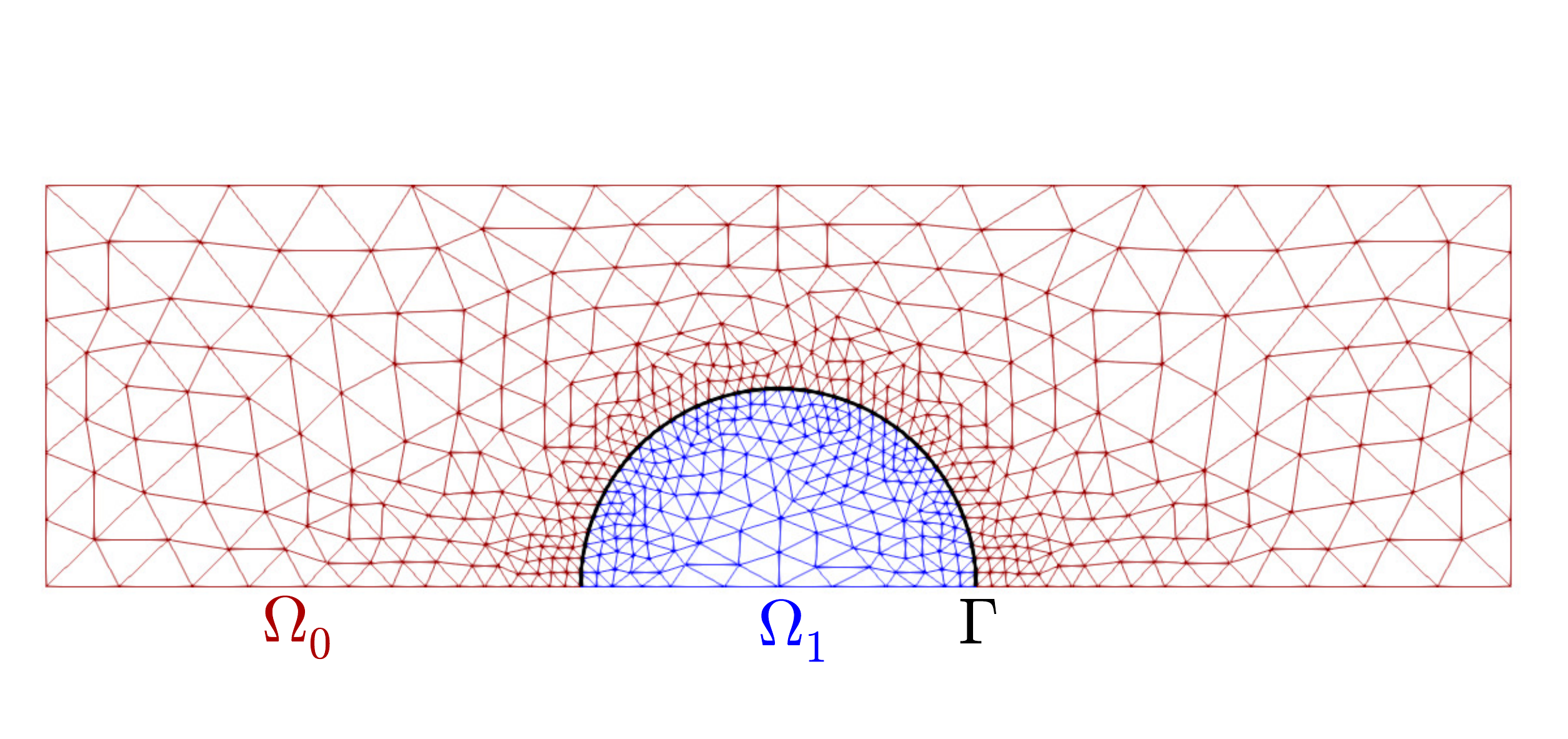}\label{fig:meshes}
\caption[]{Example of the mesh for $\Omega$ which is composed of $\Omega_0$, $\Omega_1$ and $\Gamma$. The blue colored half circular mesh refers to the fluid inside the elastic shell $\Omega_1$. The black line denotes the membrane $\Gamma$. Every grid point on the interface of $\Omega_0$ has a corresponding grid point on the interface of $\Omega_1$ sharing the same point coordinates. The lower horizontal boundary is the axisymmetry axis.}
\end{figure}

We use a Finite-Element method where the grids to represent the domains are matched at the immersed interface, i.e. they share the same grid points. Accordingly, let $T_{h,i}, i=0,1$ be the triangulations of $\Omega_i$ such that $T_h = T_{h,0}\cup T_{h,1}$ is a conforming triangulation of $\Omega$. The triangulation of the membrane is given by $\Gamma_h = T_{h,0}\cap T_{h,1}$. An example for the corresponding numerical mesh is shown in \autoref{fig:meshes}. The triangulations $T_{h,0}$ and $T_{h,1}$ are separated, i.e. $\Gamma_h$ acts as a boundary for both. 
This definition of the mesh ensures the possibility of continuous velocity and non-zero pressure jump across $\Gamma$. See the appendix for further details on the technical realization.

In order to obtain the membrane force, the tangential $\mathbf{t}$ and the values for $\mathbf{n}$, $\kappa$, $K$, $\Delta_{\Gamma}\kappa$, $\lambda_1$, and $\lambda_2$ have to be calculated. Consider the parametrization $\mathbf{X}(s)$ with the arc length parameter $s$ (hence, $\left\|\mathbf{X}_s\right\|=1$) we obtain from \cite{HuJCP2014}
\begin{align}
\mathbf{t} &= \left(X_s, R_s\right)^T, \quad &&\mathbf{n}= \left(-R_s, X_s\right)^T, \quad &&\nabla_{\Gamma}\lambda_k = \mathbf{t}\,\partial_s\lambda_k,\ \  k\in\left\{0,1\right\} \\
\kappa &= \frac{X_s}{R} + R_sX_{ss}-R_{ss}X_s, \quad &&K = \frac{X_s\left(R_sX_{ss}-R_{ss}X_s\right)}{R}, 
\quad&&\Delta_{\Gamma}\kappa = \frac{\kappa_sR_s}{R} + \kappa_{ss}.\quad
\label{eq:curvaturesAndLambdas}
\end{align}

In the discrete case, $\mathbf{X}^i = \left(X^i,R^i\right)$ for $i=0,...,N-1$ is the sequence of membrane grid points ordered counterclockwise, approximating $\Gamma$ by piece wise linear line segments $\left(\mathbf{X}^i,\mathbf{X}^{i+1}\right)$. Then, the derivatives of $X^i$, $R^i$ and $\kappa^i$ with respect to $s$ can be calculated with finite differences, here shown for $X^i$, where $i = 1,...,N-2$
\begin{align}
X_s^i = \frac{X^{i+1}-X^{i-1}}{\left\|\mathbf{X}^{i+1}-\mathbf{X}^{i-1}\right\|}\, , \qquad X_{ss}^i = \frac{2}{\left\|\mathbf{X}^{i+1}-\mathbf{X}^{i-1}\right\|}\left(\frac{X^{i+1}-X^{i}}{\left\|\mathbf{X}^{i+1}-\mathbf{X}^{i}\right\|} - \frac{X^{i}-X^{i-1}}{\left\|\mathbf{X}^{i}-\mathbf{X}^{i-1}\right\|}\right)\, .
\label{eq:centralDifferences}
\end{align}
The approximations for the principal stretches read
\begin{align}
\lambda_1^i = \frac{\left(\left\|\mathbf{X}^{i+1}-\mathbf{X}^{i}\right\|+\left\|\mathbf{X}^{i}-\mathbf{X}^{i-1}\right\|\right)}{\left(\left\|\mathbf{X}^{i+1}_{\text{ref}}-\mathbf{X}_{\text{ref}}^{i}\right\|+\left\|\mathbf{X}_{\text{ref}}^{i}-\mathbf{X}_{\text{ref}}^{i-1}\right\|\right)}, \quad \lambda_2^i = \frac{R^i}{R^i_{\text{ref}}}.\label{eq:principalStretchesDiscrete}
\end{align}
The usage of $\lambda_1^i$ to compute the derivative $\partial_s \lambda_1$ turned out to be numerically unstable. Hence, instead of the vertex-based values $\lambda_1^i$ we use the following values which represent the stretching at the midpoints of the two neighboring edges of vertex $i$
\begin{align}
\lambda_1^{i+\frac{1}{2}} = \frac{\left\|\mathbf{X}^{i+1}-\mathbf{X}^{i}\right\|}{\left\|\mathbf{X}^{i+1}_{\text{ref}}-\mathbf{X}^{i}_{\text{ref}}\right\|}, \quad 
\lambda_1^{i-\frac{1}{2}} = \frac{\left\|\mathbf{X}^{i}-\mathbf{X}^{i-1}\right\|}{\left\|\mathbf{X}^{i}_{\text{ref}}-\mathbf{X}^{i-1}_{\text{ref}}\right\|},
\end{align}
With these quantities the derivatives of the principal stretches are computed, 
\begin{align}
\partial_s\lambda_1^i = \frac{\lambda_1^{i+\frac {1}{2}}-\lambda_1^{i-\frac{1}{2}}}{\frac{1}{2}\left(\left\|\mathbf{X}^{i+1}-\mathbf{X}^{i}\right\|+\left\|\mathbf{X}^{i}-\mathbf{X}^{i-1}\right\|\right)},\quad \partial_s\lambda_2^i = \frac{\lambda_2^{i+1}-\lambda_2^{i-1}}{\left\|\mathbf{X}^{i+1}-\mathbf{X}^{i-1}\right\|}. \label{eq:ds_lambda}
\end{align}
To calculate the surface quantities on the symmetry axis (e.g. $i=0$ and $i=N-1$), one takes advantage of the fact that $R= R_{ss} = R_{ssss}=0$ and $X_s=X_{sss}=0$ and applies l'Hospital's rule, see \cite[Sec. 3.1]{HuJCP2014} for details.

\subsection{Weak form}
The fully discrete system in weak form is presented in the following. The finite element (FE) spaces read
\begin{align}
V_h &= \left\{v\in H^1_0(\Omega) \,|\, v_{|k} \in P_2(k), k\in T_h\right\} \\
M_{h,i} &= \left\{q\in L^2_0(\Omega_i)\cap C(\overline{\Omega}_i) \,|\, q_{|k} \in P_1(k), k\in T_{h,i}\right\} ,\quad i = 0,1,
\end{align}
where $V_h$ is the FE space for the velocity. The resulting velocity will be continuous across $\Gamma_h$. The FE spaces $M_{h,i}$ for the pressures $p_i$ are defined separately in $T_{h,i}$ to allow discontinuous pressure across $\Gamma_h$. Assuming constant viscosities $\eta_i$ in $T_{h,i}$, the weak form of the system given in \autoref{sec:axi} reads:
\begin{align*}
\text{Find } \left({\bf v}^n, p_0^n, p_1^n\right)\in V_h^2 \times M_{h,0} \times M_{h,1} \text{, s.t. } \forall \left({\bf w},q_0,q_1\right)\in V_h^2 \times M_{h,0} \times M_{h,1}:
\end{align*}
\begin{align}
0=&&\int_{\Omega_0(t^n)} &\rho_0\left( \frac{\mathbf{v}^n-{\mathbf{v}}^{n-1}_{\text{moved}}}{\tau} + \left(\mathbf{v}^{n-1}-\mathbf{v}^{n-1}_{\text{grid}}\right)\cdot\nabla\mathbf{v}^n\right)\cdot\mathbf{w} + \eta_0\nabla{\bf v}^n\cdot\nabla{\bf w} - p_0^n\nabla\cdot{\bf w} \nonumber\\
&&& + \left(0, \frac{\eta_0}{R}\right)\cdot\left(\nabla\mathbf{v}+ \nabla\left(\mathbf{v}\right)^T\right)\cdot\mathbf{w} +  \frac{2\eta_0}{R^2}v_r  \begin{pmatrix} 0 \\ 1 \end{pmatrix}\cdot\mathbf{w}\nonumber\\
&&+\int_{\Omega_1(t^n)} &\rho_1 \left(\frac{\mathbf{v}^n-{\mathbf{v}}^{n-1}_{\text{moved}}}{\tau} + \left(\mathbf{v}^{n-1}-\mathbf{v}^{n-1}_{\text{grid}}\right)\cdot\nabla\mathbf{v}^n\right)\cdot\mathbf{w} +\eta_1\nabla{\bf v}^n\cdot\nabla{\bf w} - p_1^n\nabla\cdot{\bf w} \nonumber\\
&&& + \left(0, \frac{\eta_1}{R}\right)\cdot\left(\nabla\mathbf{v}+ \nabla\left(\mathbf{v}\right)^T\right)\cdot\mathbf{w} +  \frac{2\eta_1}{R^2}v_r  \begin{pmatrix} 0 \\ 1 \end{pmatrix}\cdot\mathbf{w}\nonumber \\
&&- \int_{\Gamma} &
\left(\frac{\partial E_{\text{stretch}}}{\partial\Gamma}
+\frac{\partial E_{\text{tension}}}{\partial\Gamma}
+\frac{\partial E_{\text{bend}}}{\partial\Gamma}\right)^{n-1}\cdot{\bf w}\nonumber\\
0=&&\int_{\Omega_0(t)} &\tilde{\nabla}\cdot\mathbf{v}^n q_0 + \int_{\Omega_1(t)} \tilde{\nabla}\cdot\mathbf{v}^n q_1\, ,\label{eq:weakForm}
\end{align}
with surface forces $\frac{\partial E_{\cdot}}{\partial\Gamma}$ defined by Eqs. \eqref{eq:surfaceTensionForce}-\eqref{eq:ds_lambda}.

\subsection{Mesh movement}\label{sec:meshMovement}
The basic idea of the ALE approach is to move the grid of the elastic structure with the material velocity, while the fluid grid is moved with an arbitrary velocity keeping the mesh in a proper shape. 
Accordingly, in the present work the membrane grid points are displaced with the velocity $\mathbf{v}$ in every time step. As the membrane is moving, it is necessary to extend this movement to every grid point in $T_h$ to keep the mesh well in shape. Two different strategies for rearranging grid points on a mesh with given boundary movement have been used for the simulations in the present work.

\subsubsection{Solving Laplace problem}\label{sec:laplaceProblem}   
The first strategy involves the harmonic extension of interface movement by solving the Laplace problem
\begin{align}
\Delta\mathbf{v}_{\text{grid}} &= 0 & \mbox{in}~\Omega, \nonumber\\
\mathbf{v}_{\text{grid}}&=0 & \mbox{on}~\delta\Omega\backslash\Gamma, \nonumber\\
\mathbf{v}_{\text{grid}}&= \mathbf{v}_{} & \mbox{on}~\Gamma, \label{eq:laplaceProb}
\end{align}
where $\mathbf{v}_{\text{grid}}$ is the velocity of the grid points in $T_h$. The mesh is then moved with $\mathbf{v}_{\text{grid}}$. In some cases, it can be helpful to use the Bilaplace problem instead:
\begin{align}
\Delta^2\mathbf{v}_{\text{grid}} &= 0 & \mbox{in}~\Omega, \nonumber\\
\Delta\mathbf{v}_{\text{grid}} &= \mathbf{v}_{\text{grid}} = 0 & \mbox{on}~\delta\Omega\backslash\Gamma, \nonumber\\
\mathbf{v}_{\text{grid}}&= {\bf v}\, , \quad \mathbf{n}\cdot\nabla\Delta{\mathbf{v}_{\text{grid}}}= 0 & \mbox{on}~\Gamma. \label{eq:bilaplaceProb}
\end{align}
Both approaches can be used in most cases. However, for strong and/or periodic deformations, some problems can occur using the Laplace Problem for mesh smoothing, e.g. elements near the interface can degenerate slowly and cause a crash of the simulation. In this case, the mesh smoothing approach shown in the following may be a better choice.

\subsubsection{Element area and length conservation}\label{sec:areaAndLengthConservation} 

The second strategy is motivated by the fact that solving the Laplace problem \autoref{eq:laplaceProb} or the Bilaplace problem \autoref{eq:bilaplaceProb} can lead to degenerated elements. To avoid this, it is helpful to penalize changes in area and edge length of the elements in order to prevent large deformations in both, the area and the aspect ratio of the elements. How to preserve element surface areas or grid point distances is described in \cite{teschner2004}. 

\begin{figure}
\centering
\includegraphics[width=0.6\textwidth]{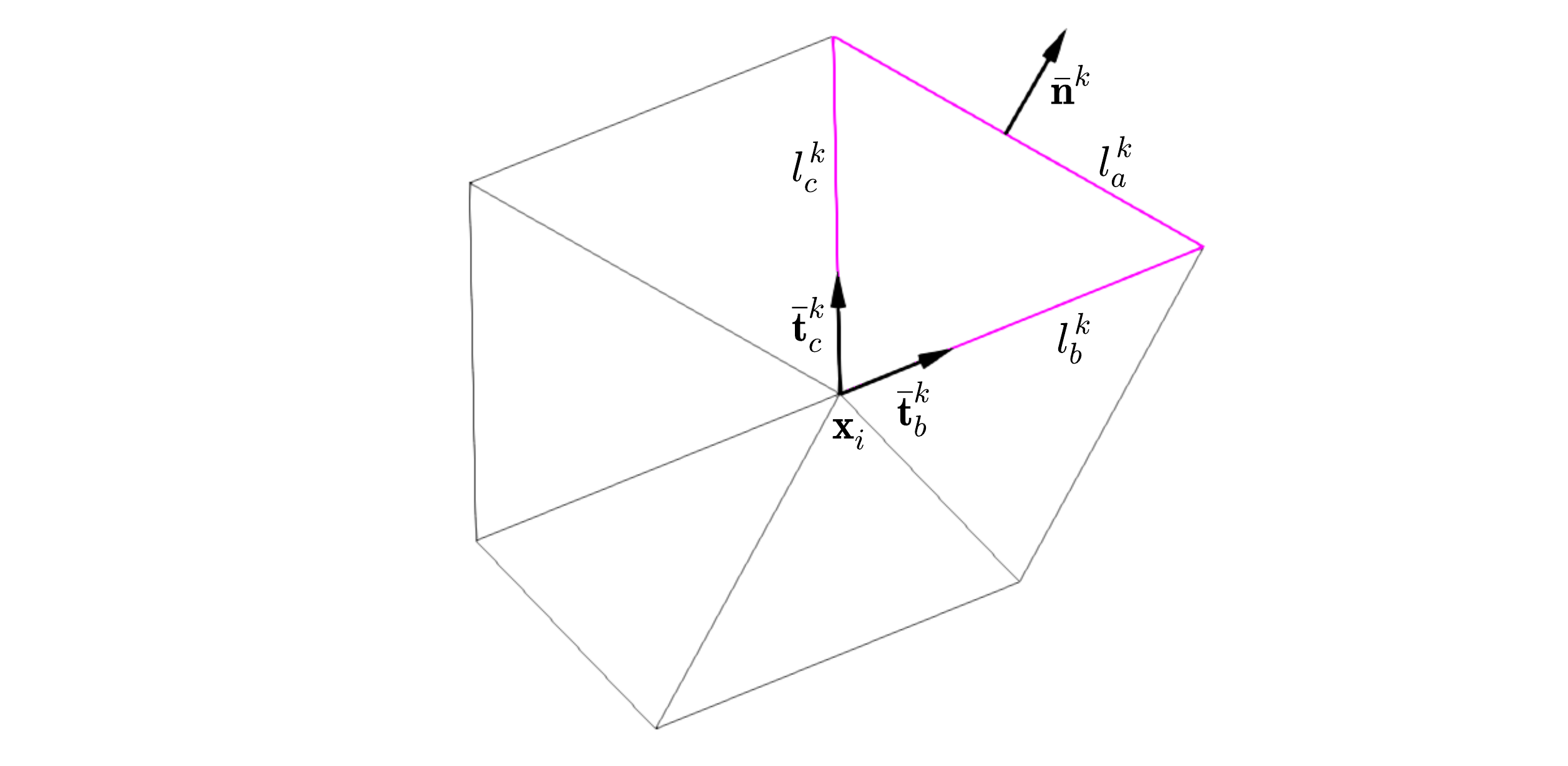}\label{fig:meshSmoothing}
\caption[]{Illustration for the mesh smoothing scheme. The point $\mathbf{x}_i$ is the point to be moved using \autoref{eq:MeshSmoothing}. A displacement value is calculated penalizing differences in areas and edge lengths of every element containing the point $\mathbf{x}_i$. }
\end{figure}

Assume that the boundary has been moved already. Iterate over all grid points $\mathbf{x}_i$. Check whether the length of each edge and/or the area of each element, which has $\mathbf{x}_i$ as a vertex, has changed. If, for example, the area of an element, which has the vertex $\mathbf{x}_i$, decreased after the movement of $\Gamma$, $\mathbf{x}_i$ will be moved in order to re-increase the area of this element. Since this has to be done for all elements that share $\mathbf{x}_i$ as a vertex, a mean value of the calculated displacements is computed. The scheme for calculating the grid point movement in order to get the new grid reads:
\begin{enumerate}
\item Calculate areas $A$ and edge lengths $l_a$, $l_b$, $l_c$ of all elements using the point coordinates.
\item Compute the velocity of every grid point $\mathbf{x}_i$ on $T_h /(\partial T_h\cup\Gamma_h)$, using the following formula
\begin{align}
{\bf v}_{\rm grid}^i = \frac{1}{\tau} \sum_{k\in N(i)} \left(c_{a}\left(1-\frac{A^k_{\text{ref}}}{A^k}\right)\bar{\mathbf{n}}^k + c_{l}\left(1-\frac{\left(l^k_b\right)_{\text{ref}}}{l^k_b}\right)\bar{\mathbf{t}}^k_b + c_{l}\left(1-\frac{\left(l^k_c\right)_{\text{ref}}}{l^k_c}\right)\bar{\mathbf{t}}^k_c 
\right), \label{eq:MeshSmoothing}
\end{align}
where $N(i) = \{e\in T_h\, |\, \mathbf{x}_i$ is a vertex of $e\}$ is the set of all elements that share the vertex $\mathbf{x}_i$, $A^k$ is the area of the element $k$, $\bar{\mathbf{n}}^k$ is the normal of the opposite edge of $\mathbf{x}_i$ in the $k$-th neighbour, $\bar{\mathbf{t}}^k_b$ and $\bar{\mathbf{t}}^k_c$ are the tangential vectors of the edges that share the vertex $\mathbf{x}_i$ pointing away from $\mathbf{x}_i$ in element $k$, and $c_a$ and $c_l$ are constants, respectively. See also \autoref{fig:meshSmoothing} for a visualization of the quantities introduced in \autoref{eq:MeshSmoothing}.  
\item Move $i$-th mesh point by $\tau{\bf v}_{\rm grid}^i$.
\end{enumerate}

\begin{figure}
\centering
\subfigure[initial mesh]{\includegraphics[width=0.3\textwidth]{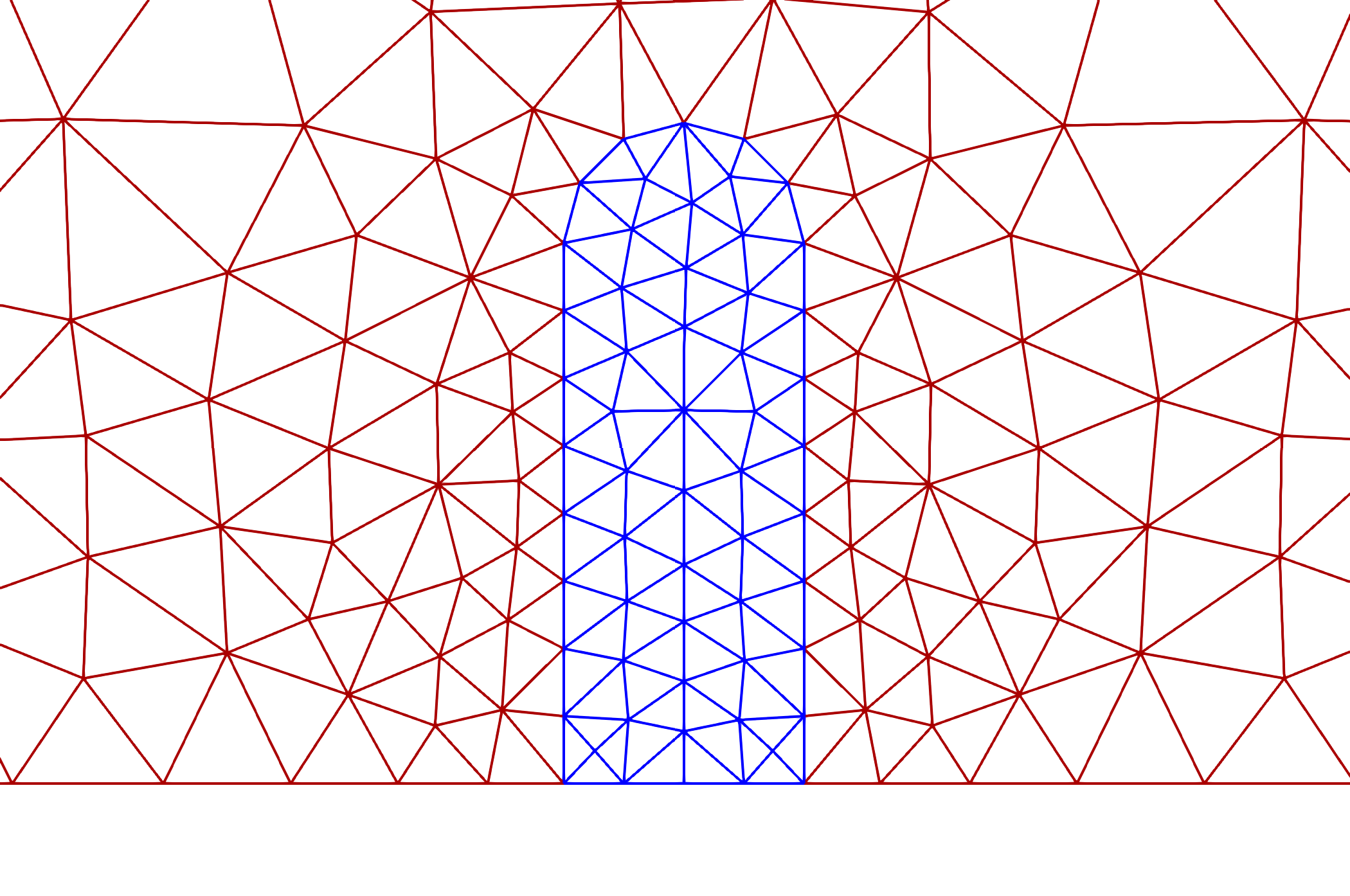}}\hspace{0.1cm}
\subfigure[Laplace]{\includegraphics[width=0.3\textwidth]{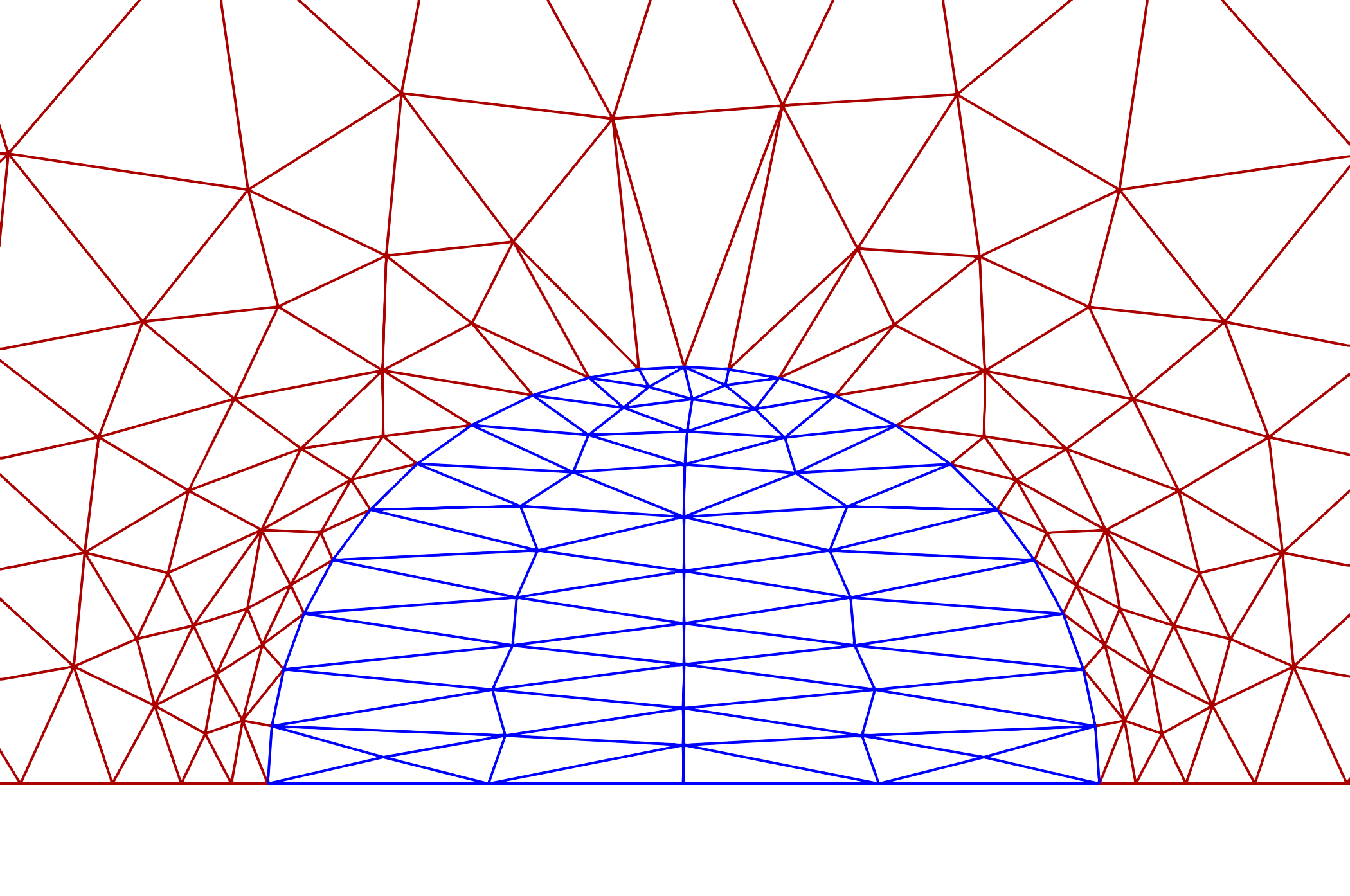}
\label{fig:meshSmoothingResults-b}}
\subfigure[area/length conservation]{\includegraphics[width=0.3\textwidth]{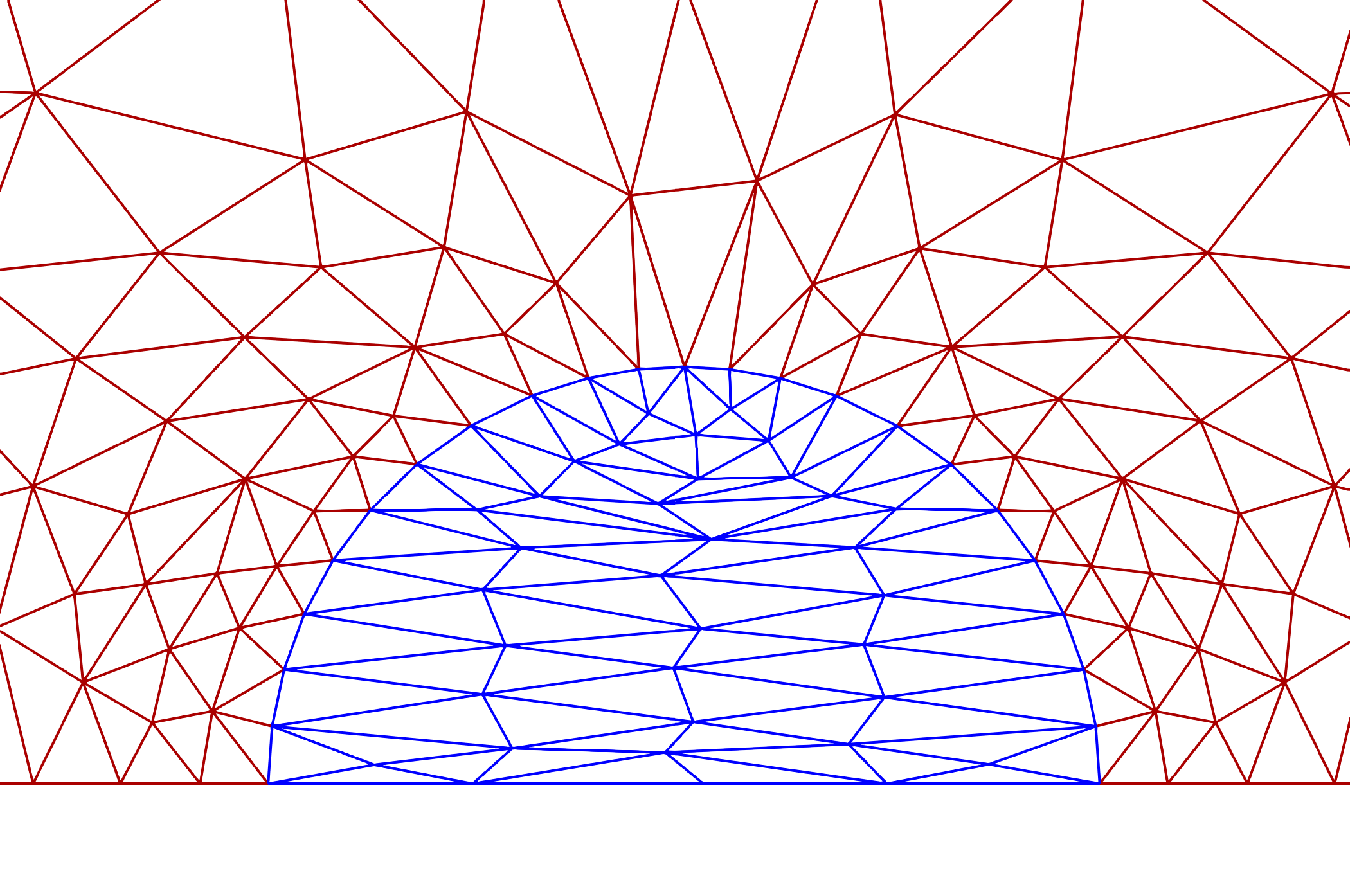}
\label{fig:meshSmoothingResults-c}}
\label{fig:meshSmoothingResults}
\caption[]{Example images for the two mesh smoothing approaches. (a) shows the mesh near the membrane before the deformation. (b) and (c) show the mesh after a strong deformation during 500 time steps. (b) is the result of the Laplace smoothing approach, where (c) is the result of the area and length conservation approach.}
\end{figure}

\autoref{fig:meshSmoothingResults} shows an example for both, the Laplace smoothing approach and the area and length conservation approach. In both cases, a strong deformation has been imposed within a total amount of 500 time steps. The membrane shape at the end is circular with perfect agreement in both cases. The Laplace smoothing result (\autoref{fig:meshSmoothingResults-b}) seems a little more even in the internal fluid. However, the area and length conservation approach (\autoref{fig:meshSmoothingResults-c}) produces a better result: In the external fluid, the triangles near the membrane are less deformed. Around the symmetry axis, elements have been less compressed in $x$ direction where at the poles, they have been less stretched in $r$ direction (the latter is also visible in the internal fluid). These rather small improvements of the second approach can be quite important, e.g. when due to a periodic deformation small errors that occur in the Laplace smoothing accumulate over time. Nevertheless, in this paper, if not mentioned otherwise, the Laplace smoothing approach is used as it is independent of additional problem-specific parameters (like $c_a, c_l$). 

\section{Numerical Tests}\label{sec:NumericalTests}

Test case simulations were performed to show the quality and capability of the  presented  model, to verify the correctness of the interfacial forces and to analyze the mesh and time step stability. 

\subsection{Verification of the interfacial forces}

\begin{figure}
\centering
\subfigure[]{\includegraphics[width=0.22\textwidth]{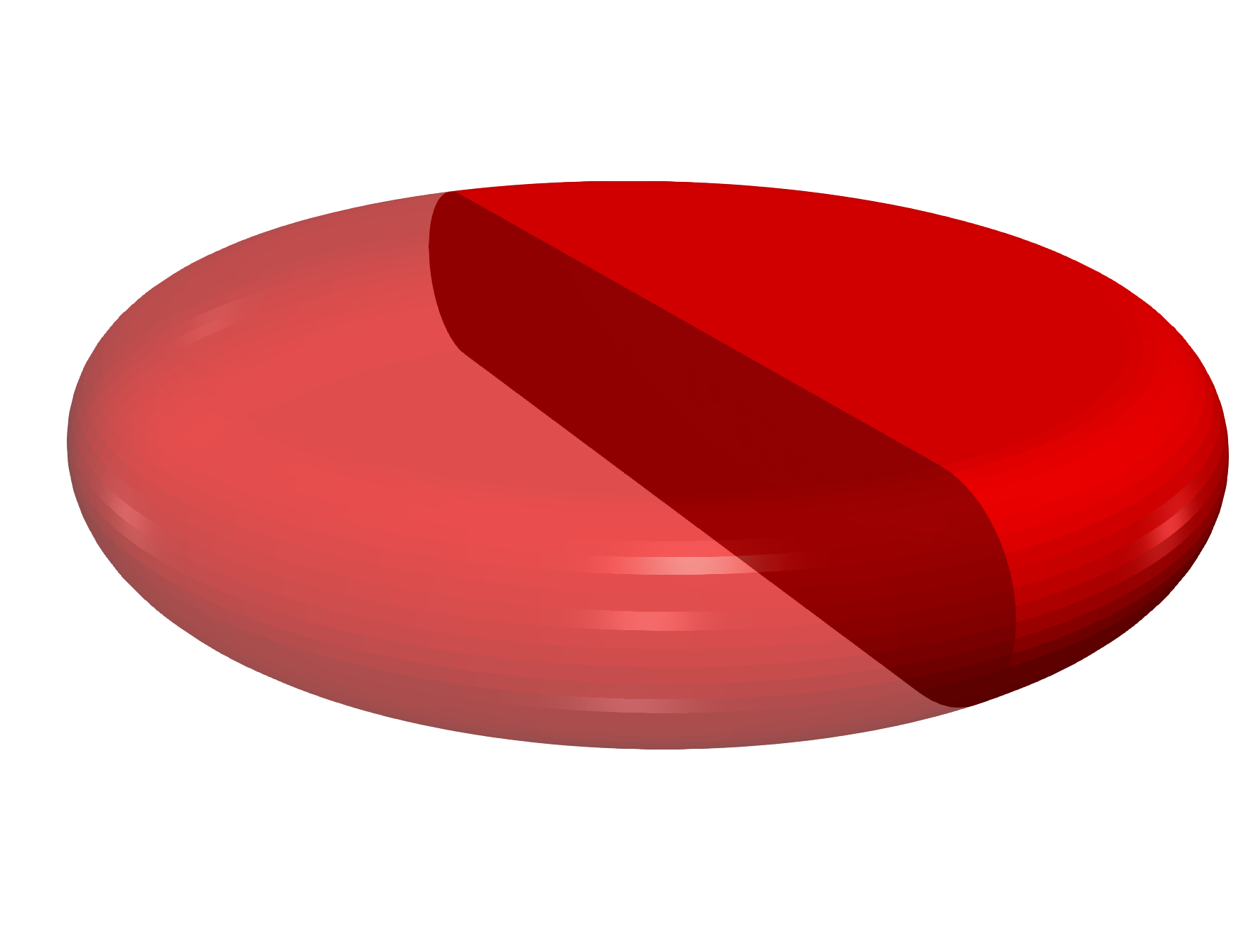}\label{fig:shellShapes-a}}
\hspace{8px}
\subfigure[]{\includegraphics[width=0.22\textwidth]{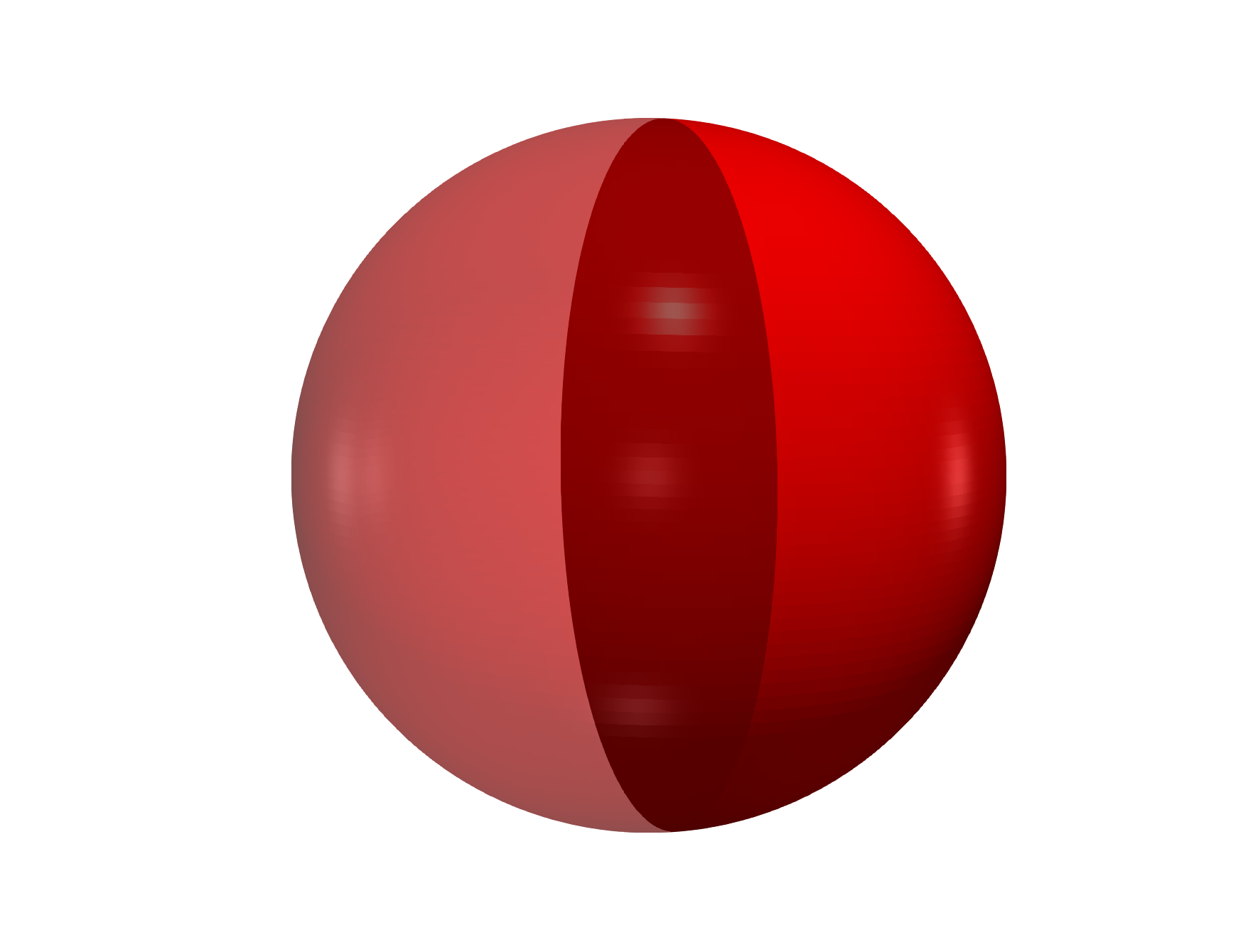}\label{fig:shellShapes-b}}
\hspace{8px}\subfigure[]{\includegraphics[width=0.22\textwidth]{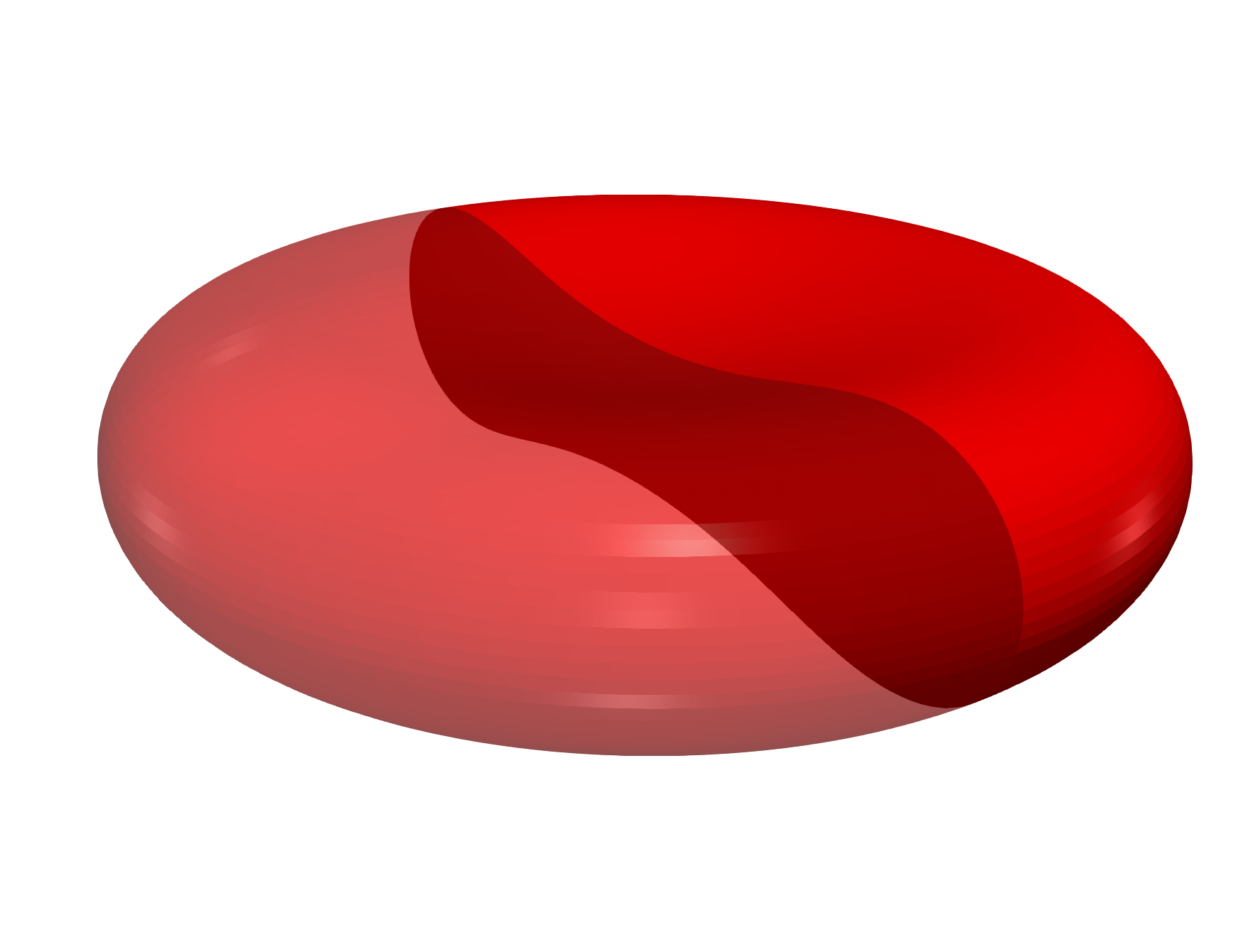}\label{fig:shellShapes-c}}
\hspace{8px}\subfigure[]{\raisebox{12px}{\includegraphics[width=0.22\textwidth]{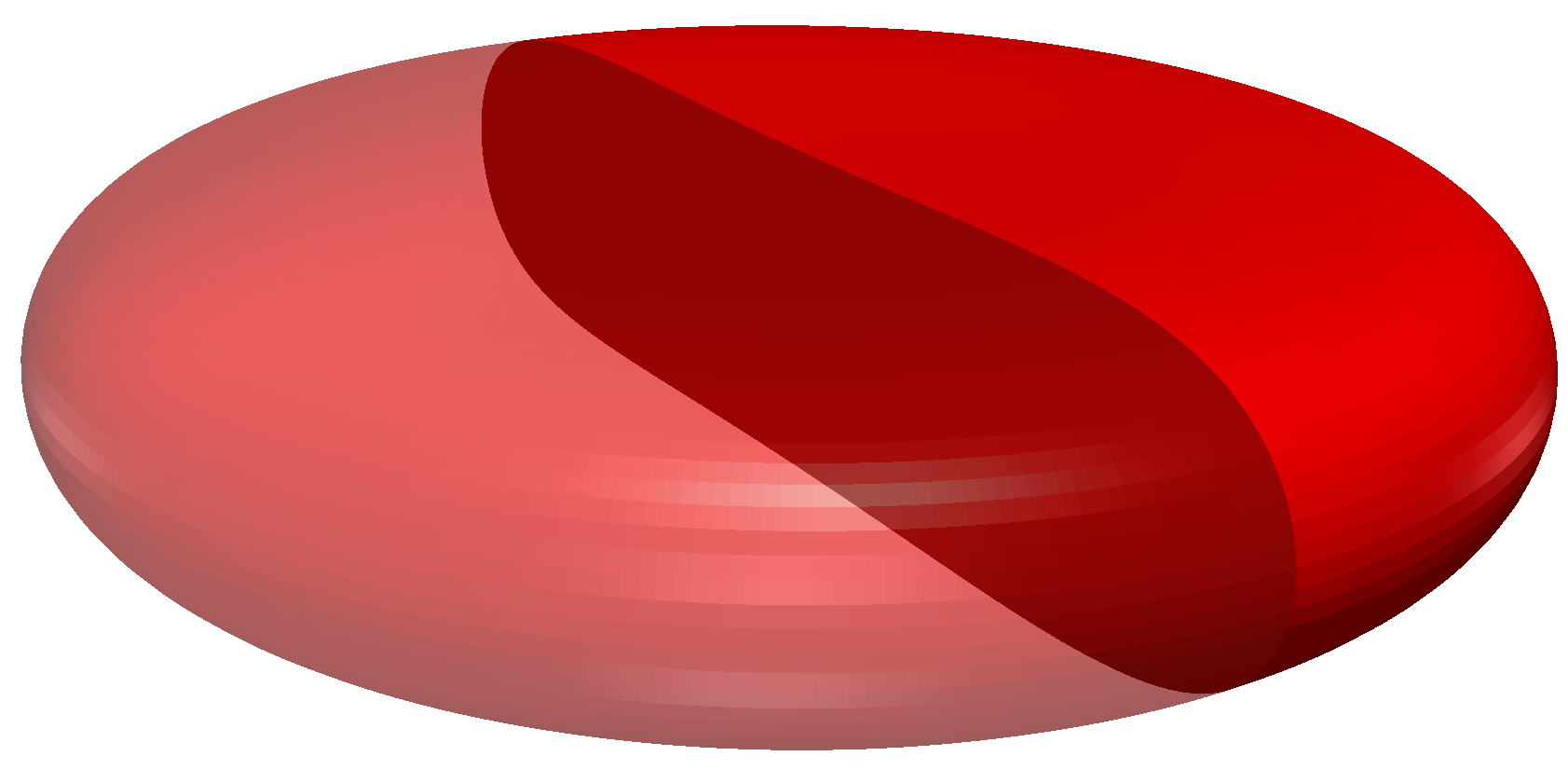}\label{fig:shellShapes-d}}}
\label{fig:shellShapes}
\caption[]{Shell shapes for the respective dominant forces on the membrane. (a) Initial shape for all cases. (b) Surface tension dominant case. Shape in stationary state. (c) Bending stiffness dominant case. Shape in stationary state. (d) Stretching dominant case. Shape in (quasi-) stationary state. All images are partly transparent to show the cross section of the elastic shell. The expected theoretical shapes are well recovered.}
\end{figure}

\begin{figure}
\centering
\subfigure[]{\includegraphics[width=0.40\textwidth]{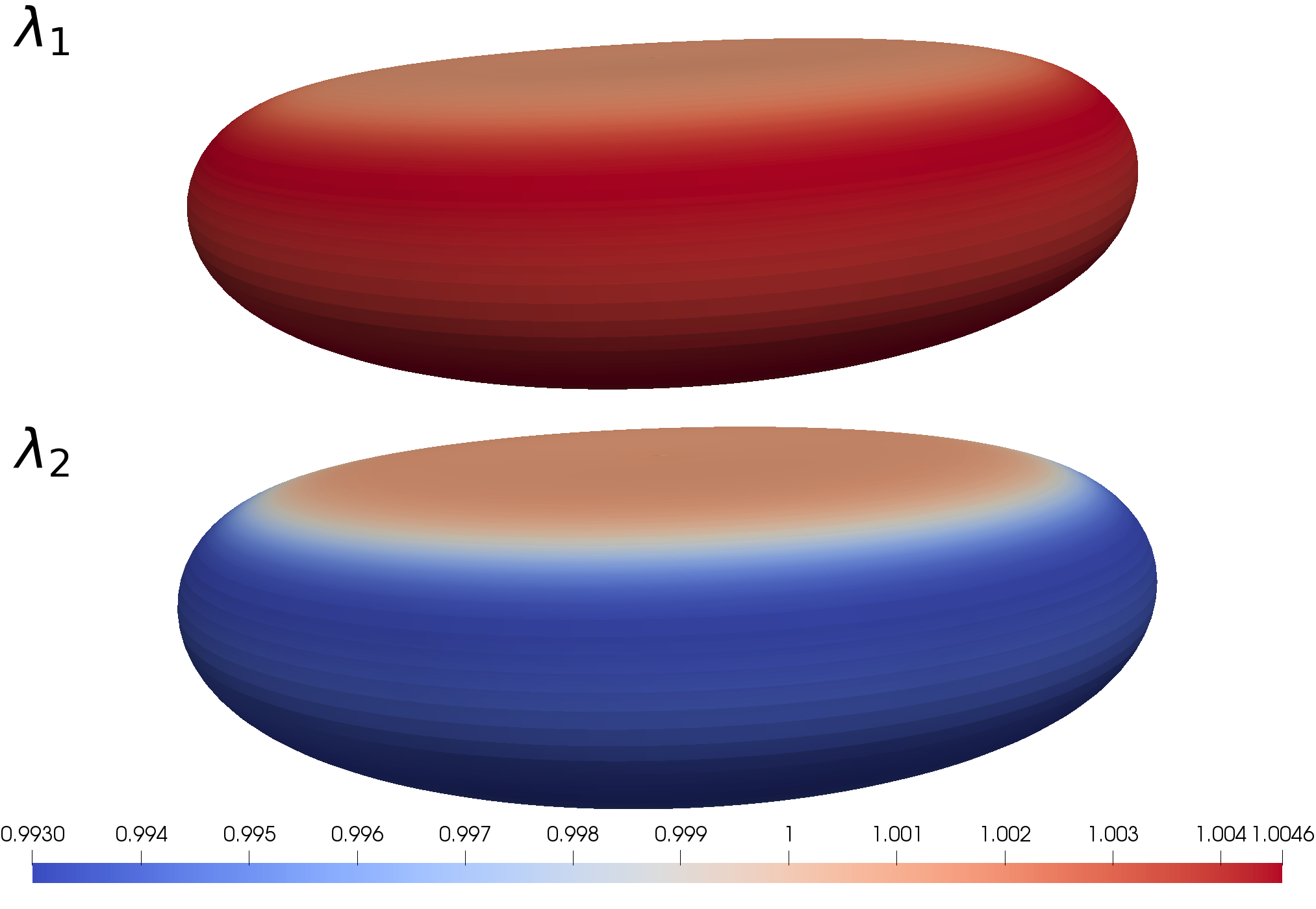}\label{fig:lambda1lambda2}}
\hspace{5px}\subfigure[]{\includegraphics[width=0.49\textwidth]{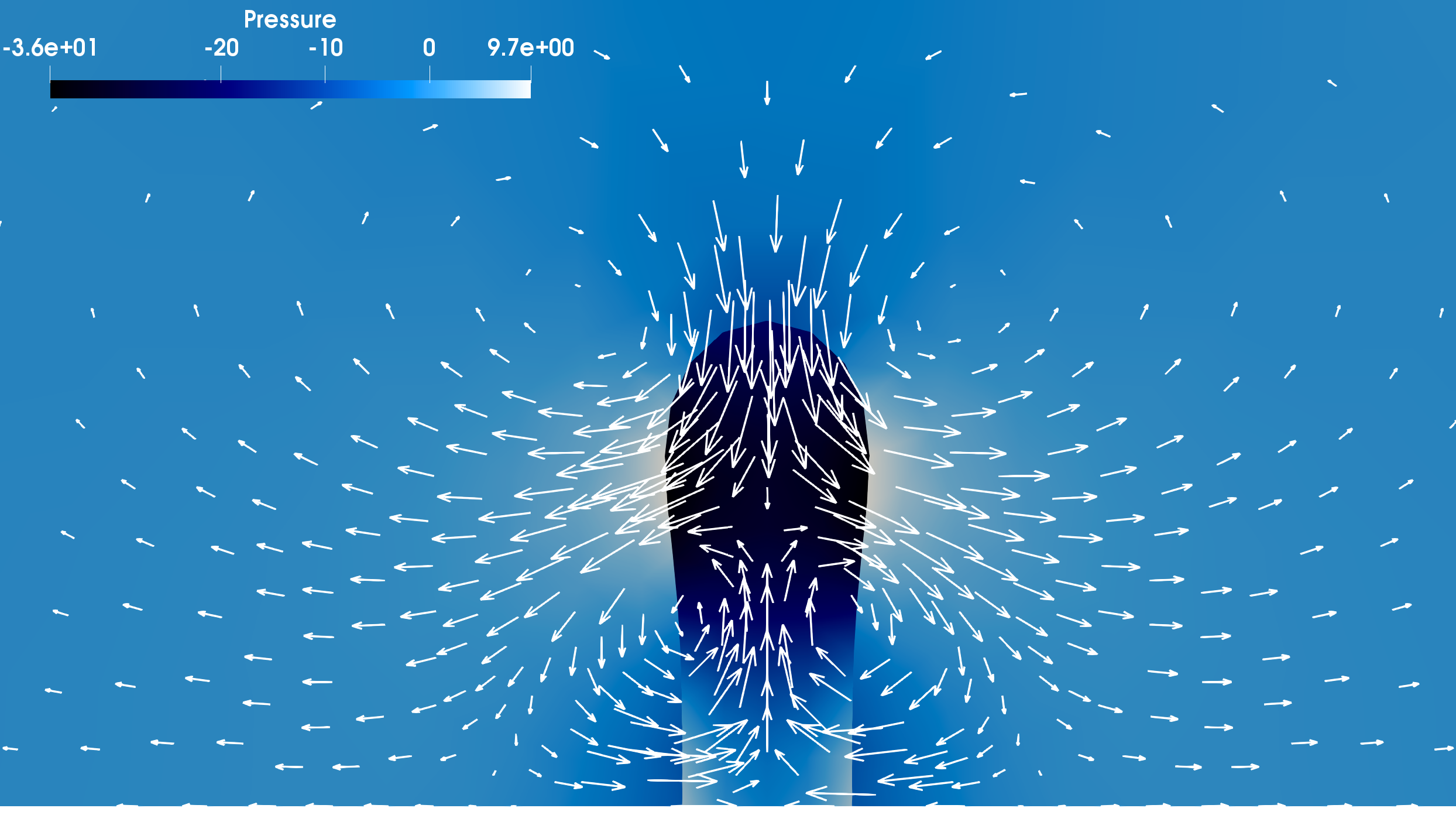}\label{fig:velocityAndPressureField}}
\caption[]{(a) The two principal stretches in the stationary state of case 3. (b) Velocity field (illustrated with vectors) and pressure field for the bending stiffness dominant case within the deformation process. As expected, the pressure discontinuous.}
\end{figure}

In the following, we prescribe the elastic membrane as an initially oblate-shaped object, i.e. its cross-section as the combination of two parallel lines and a semicircle (\autoref{fig:shellShapes-a}). The radius of the semicircle amounts to $0.1$, the parallel lines have a length of $0.9$, s.t. the surface of revolution has a equatorial radius of $0.55$ and a height of $0.2$. The membrane is located in the center of the computational domain  $\overline{\Omega}=\left[-2,2\right]\times\left[0,1\right]$. \\
To verify the correctness of the three main interfacial forces, we consider the following three configurations:
\begin{enumerate}
\item \textbf{Surface tension dominant case}: $\gamma=0.003$, $c_b=0$, $K_A=0$, $K_S=0$
\item \textbf{Bending stiffness dominant case}: $\gamma=0$, $c_b=11.1$, $K_A=0.025$, $K_S=0$
\item \textbf{Stretching dominant case}: $\gamma=0$, $c_b=0$, $K_A=0.025$, $K_S=0.008333$. Here, the initial conditions for the principal stretches $\lambda_1$ and $\lambda_2$ are changed from $1$ to $1.05$ to induce a $5\%$ pre stretch to the membrane. This causes non-zero stretching energy on the membrane.      
\end{enumerate} 
The viscosities and densities are chosen equally in both phases and amount to $\eta_i = 1$, $\rho_i = 10^3$.
\begin{figure}
\centering
\includegraphics[width=0.33\textwidth, angle=-90]{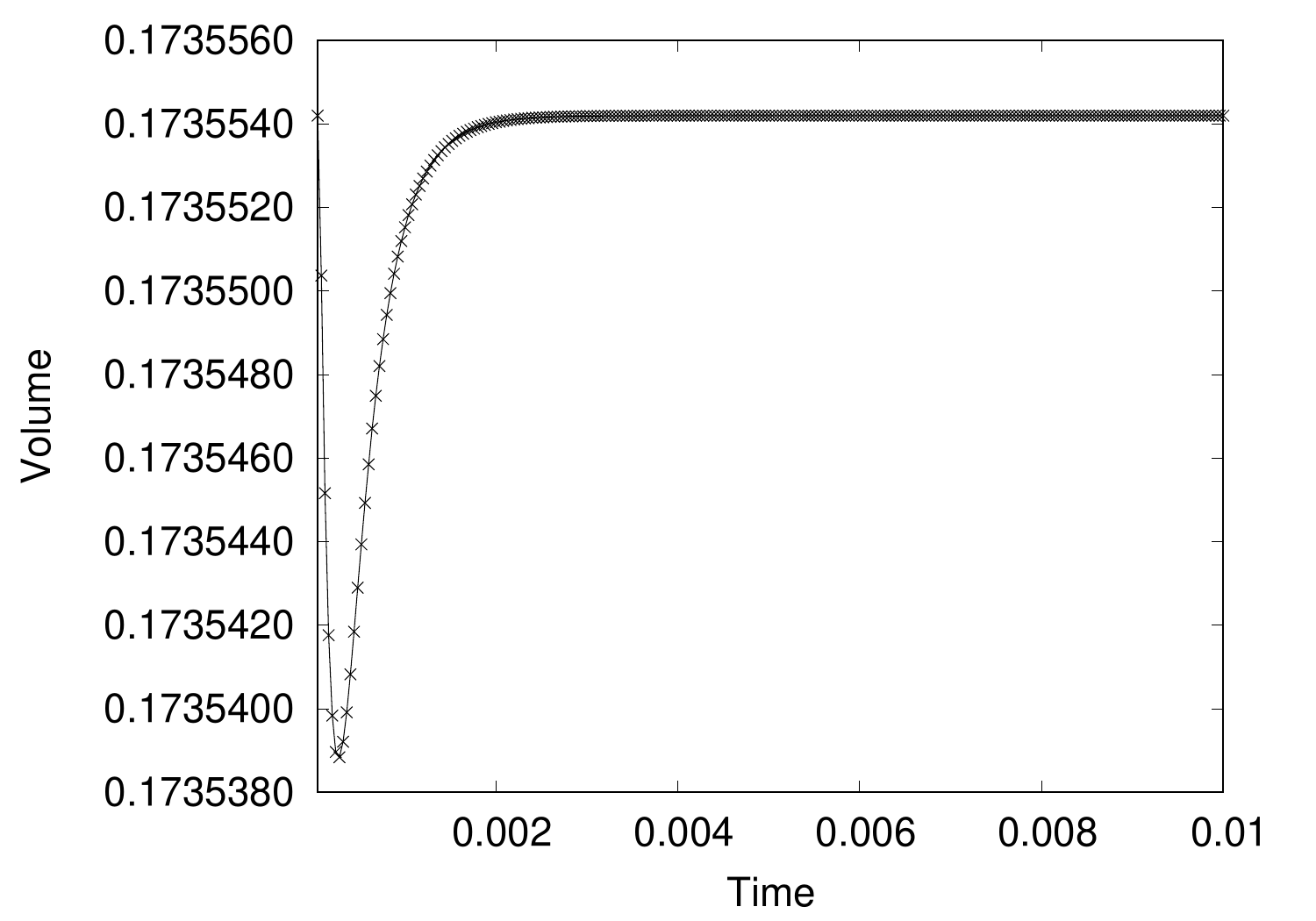}\label{fig:volumeConservation-a}
\label{fig:volumeConservation}
\caption[]{Shell volume over time for the stretching dominant case.  }
\end{figure}
\autoref{fig:volumeConservation} shows that volume conservation is perfectly ensured in the simulations. The figure shows the volume over time for the stretching dominant case. The volume decreases in the beginning for a maximum amount of $0.085\%$. The other cases show similar behaviour with even smaller volume changes in the beginning of the deformation. \\
In case 1, the membrane is expected to evolve into a sphere, since surface tension tends to minimize the surface area. This behavior is verified in the simulations, see \autoref{fig:shellShapes-b}. \\
In case 2 the stationary state is expected to yield a red blood cell like shape, since bending stiffness tends to minimize the curvature locally while the additional influence of area dilation prevents the membrane to deform (or stretch) strong enough to get spherical. In this sense, case 2 is similar to a lipid vesicle, as the finite $K_A$ leads to approximate conservation of surface area.
The stationary shapes of these simulations fit qualitatively well with theory \autoref{fig:shellShapes-c} \cite{HuJCP2014}. \\
In case 3, in-plane elasticity penalizes stretching in tangential direction to the surface. Hence, the initial condition of $\lambda_1$ and $\lambda_2$ being larger than $1.0$ causes a deformation of the membrane such that the radius of the oblate should decrease, where the thickness should somehow increase a bit. Note, that in the absence of volume conservation, the membrane would contract in both directions such that the principal stretches would approach $1.0$ everywhere on the membrane. However, the conservation of enclosed volume prevents the principal stretches from reaching $1.0$, in general. The stationary state of the stretching dominant case is shown in  (\autoref{fig:shellShapes-c}). The principal stretches in the stationary state are illustrated in \autoref{fig:lambda1lambda2}.  The equilibrium configuration of the elastic surface shows a significant stretch ($\lambda_1 > 1$) in lateral direction, which is necessary to accommodate the excess volume. This is accompanied by a compression in circumferential direction ($\lambda_2 < 1$) to minimize surface dilation.

In the numerical model, the Navier-Stokes equations are solved with separately defined pressures in $\Omega_0$ and $\Omega_1$, which leads to a discontinuous pressure field along $\Gamma$. The pressure field together with velocity glyphs is shown in \autoref{fig:velocityAndPressureField} for the bending stiffness dominant case.

\subsection{Mesh resolution study}
\begin{figure}
\centering
\subfigure[base mesh, $h_1$]{\includegraphics[width=0.3\textwidth, angle=0]{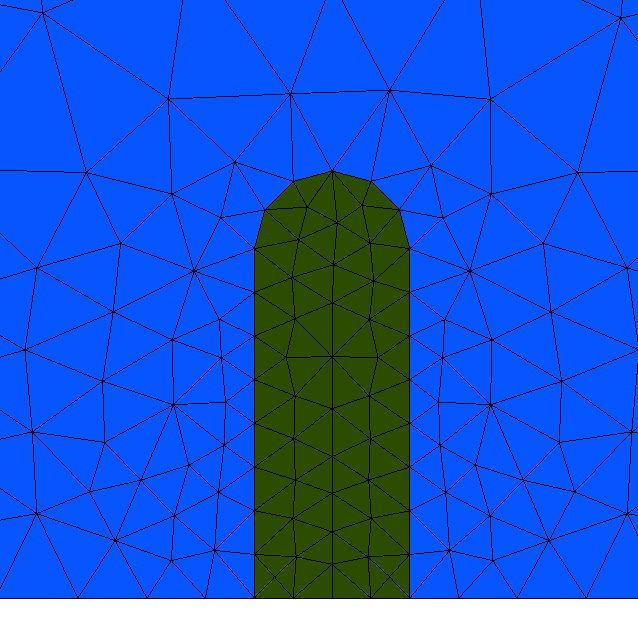}}
\subfigure[one refinement step, $h_2$]{\includegraphics[width=0.3\textwidth, angle=0]{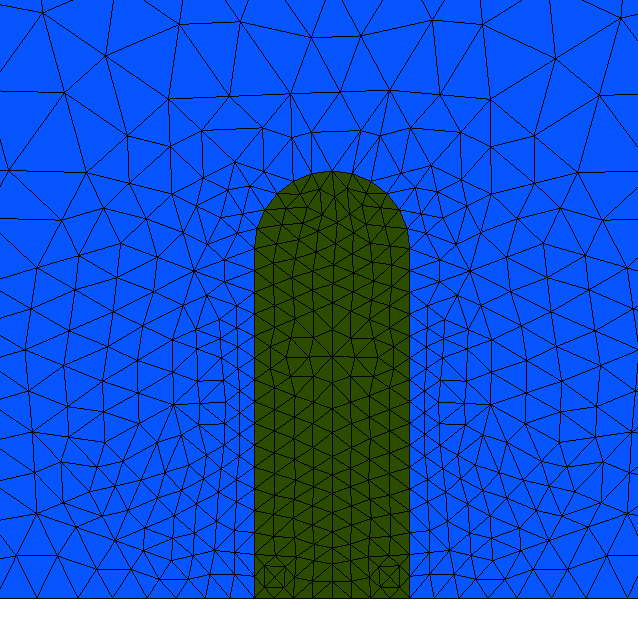}}
\subfigure[two refinement steps, $h_3$]{\includegraphics[width=0.3\textwidth, angle=0]{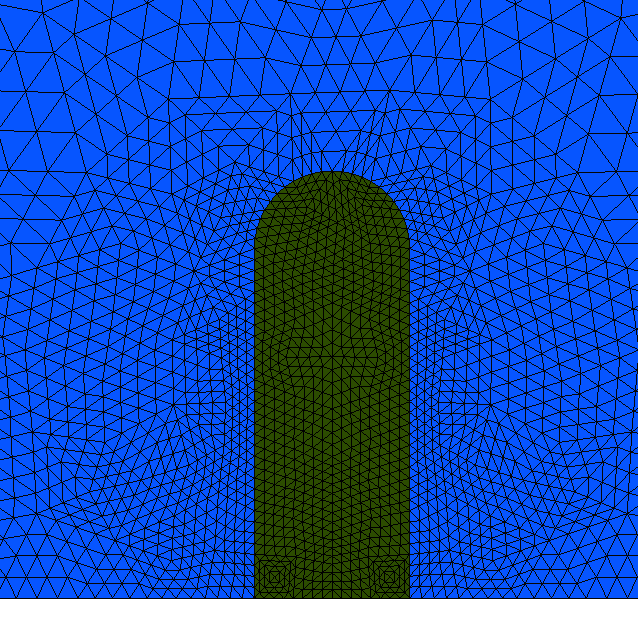}}
\label{fig:MeshCloseUps}
\caption[]{The three (initial) meshes used for the mesh resolution study. Close up views around the interface. Blue refers to $\Omega_0$, green refers to $\Omega_1$.}
\end{figure}

Three different mesh resolutions have been chosen to investigate the dependence of the simulation results on the mesh (\autoref{fig:MeshCloseUps}). In the following we denote the mesh size by $h_i$ and number of surface grid points by $N_i$ for $i\in\{1,2,3\}$. The coarsest mesh has a mesh size of $h_1=0.055$ at the interface, hence the membrane is resolved by $N_1=23$ grid points. The complete mesh has $228$ grid points. Refining this mesh by two triangle bisections leads to an intermediate mesh ($h_2=0.0275, N_2=45$, $820$ total grid points). Two further bisections lead to the finest mesh ($h_3=0.01375, N_3=89$, $3102$ total grid points).
 
\begin{figure}
\centering
\subfigure[]{\includegraphics[width=0.18\textwidth, angle=0]{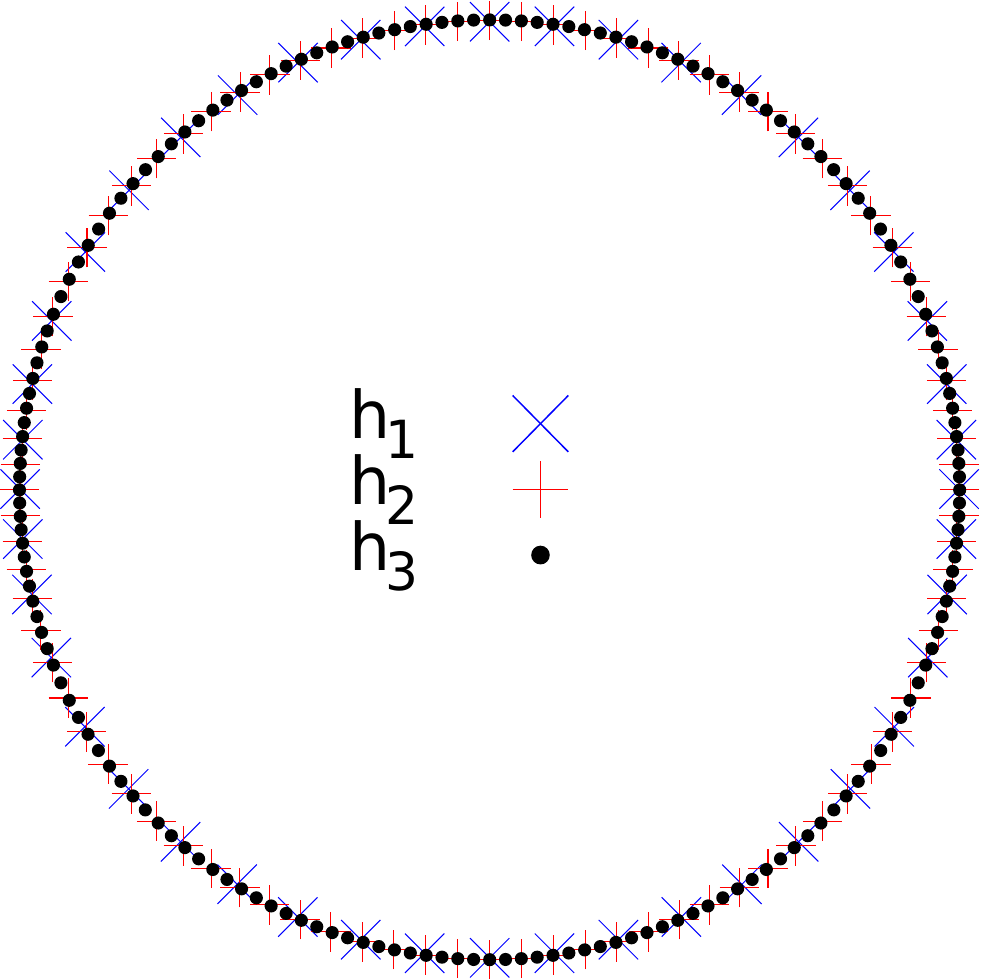}\label{fig:MeshCrossSections-a}}\hspace{0.5cm}
\subfigure[]{\includegraphics[width=0.33\textwidth, angle=0]{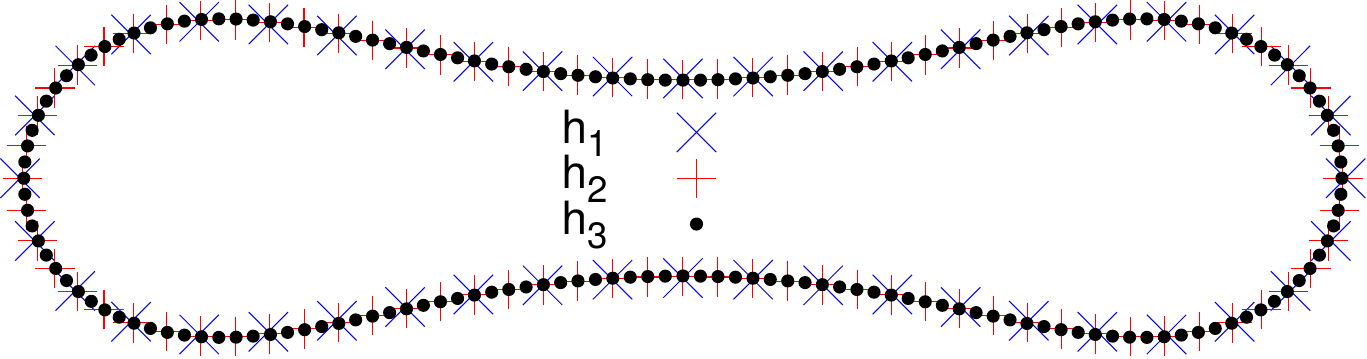}}\hspace{0.5cm}
\raisebox{7px}{\parbox{0.33\textwidth}{\subfigure[]{\includegraphics[width=0.33\textwidth, angle=0]{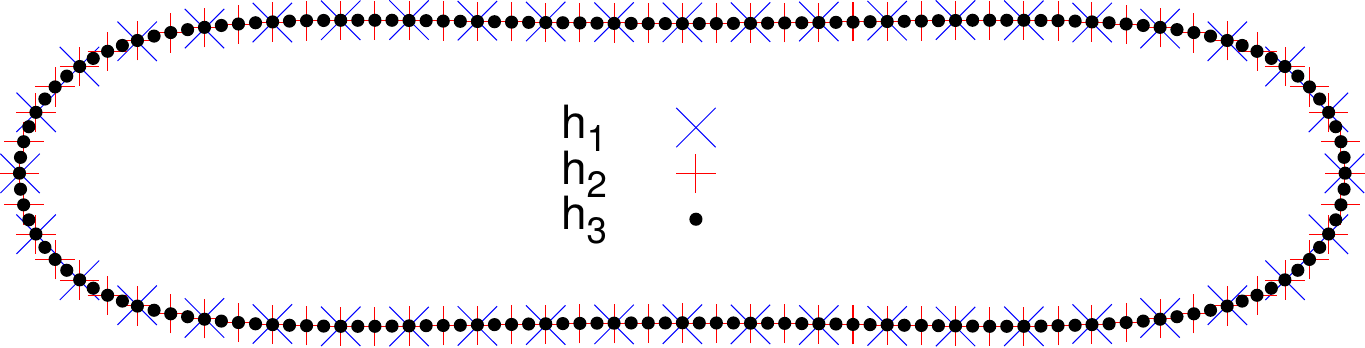}}}}
\label{fig:MeshCrossSections}
\caption[]{Shell cross section shapes for the three different cases. The three different mesh resolutions with mesh size $h_1, h_2,$ and $ h_3$ are used in each case. (a) Surface tension dominant case for t=0.02. (b) Bending stiffness dominant case for t=0.0006. (c) Stretching dominant case for t=0.01.}
\end{figure}

\autoref{fig:MeshCrossSections} shows the membrane points of the respective case for all chosen mesh resolutions. As can be seen, the method produces accurate results even with relatively coarse meshes. There is no visible difference between the shapes of all three meshes. To quantify that, we introduce two different error measures in the following. 

The mean distance error between membrane points on the $h_1$- and $h_2$-mesh and the corresponding points on the next finer mesh is defined as
\begin{align}
E^{h_i} = \frac{1}{N_1}\sum_{j=0}^{N_1-1} \left|\left|\mathbf{X}_{2^{i-1}\cdot j}^{h_i}-\mathbf{X}_{2^{i}\cdot j}^{h_{i+1}}\right|\right|\, , i \in \left\{1,2\right\} \label{eq:errorPointwiseDeviation}
\end{align}
For the $h_2$ (or $h_3$) mesh, every other (or 4th) membrane point is used, s.t. only corresponding grid points (existing in the coarsest mesh) are compared.

As a second error indicator we use the membrane cross section perimeter. Given that membrane points are sorted, we define the perimeter 
\begin{align}
P^{h_i} = \sum_{j=0}^{N_i-1} \left|\left|\mathbf{X}_j^{h_i}-\mathbf{X}_{j+1}^{h_i}\right|\right|\, , i \in \left\{1,2,3\right\} 
\end{align}
where $\mathbf{X}_{N_i}^{h_i}:=\mathbf{X}_0^{h_i}$. Then, the error is calculated using 
\begin{align}
E_{P}^{h_i} = \left|P^{h_i}-P^{h_{i+1}}\right|\, , i \in \left\{1,2\right\}. 
\end{align} 
 With these values, the experimental order of convergence (EOC$_E$ and EOC$_P$) can be obtained
\begin{align}
\text{EOC}_E = \frac{\text{ln} \frac{E^{h_1}}{E^{h_2}}}{\text{ln}\frac{h_1}{h_2}}\, , \quad
\text{EOC}_P = \frac{\text{ln}\frac{E_P^{h_1}}{E_P^{h_2}}}{\text{ln}\frac{h_1}{h_2}}.
\end{align}
The obtained values are shown in \autoref{tab:EOC}. The point coordinates converge with order 1 and the areas/perimeters converge with order 2.

\subsection{Time step study}
\begin{figure}
\centering
\subfigure[t=0.002]{\includegraphics[width=0.24\textwidth]{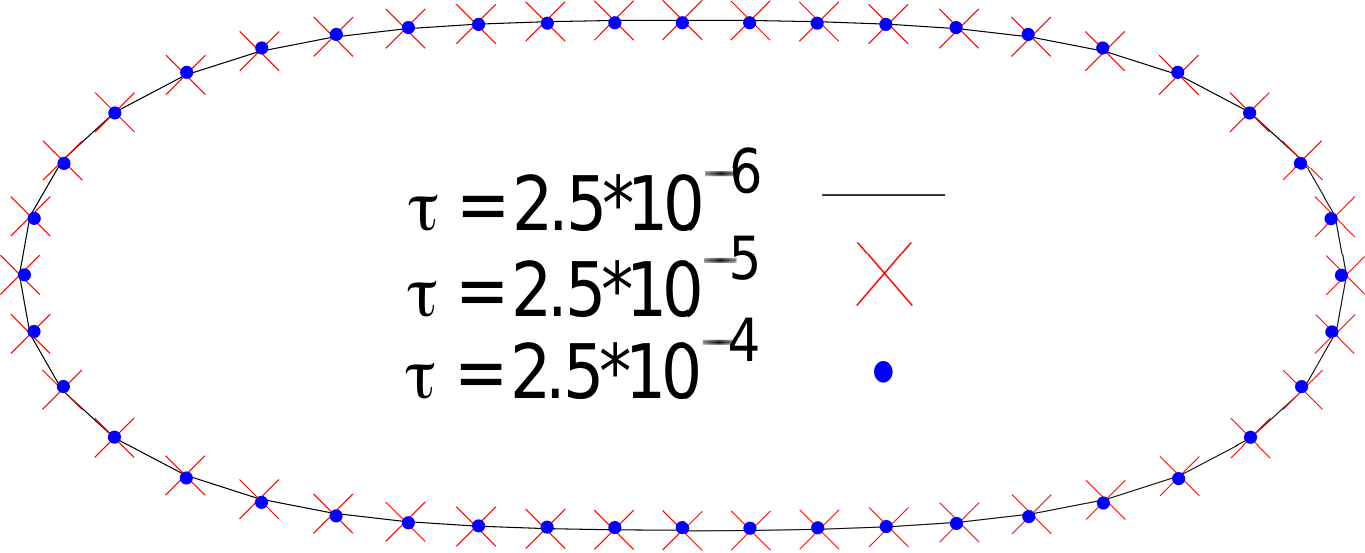}\label{fig:evolutionOfShapes-a}}\hspace{35px}
\subfigure[t=0.005]{\includegraphics[width=0.19\textwidth]{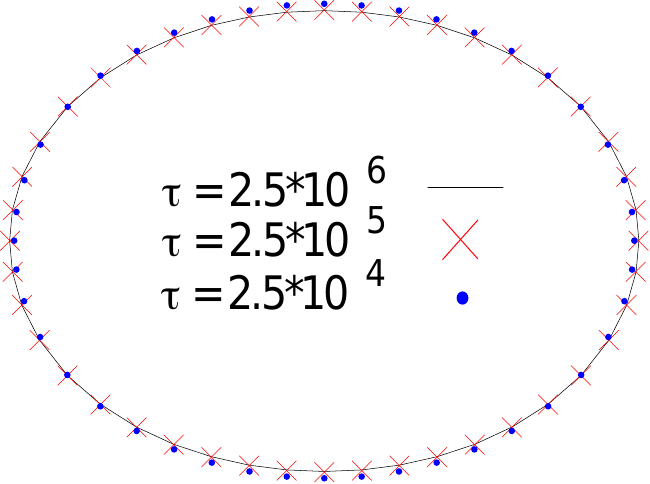}\label{fig:evolutionOfShapes-b}}\hspace{45px}
\subfigure[t=0.02]{\includegraphics[width=0.17\textwidth]{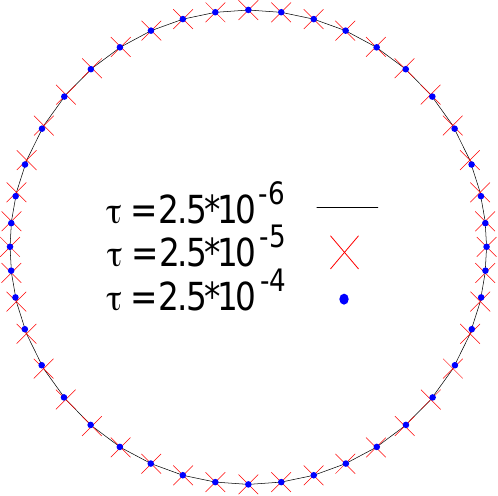}\label{fig:evolutionOfShapes-c}}\hspace{15px}

\subfigure[t=0.0005]{\includegraphics[width=0.3\textwidth]{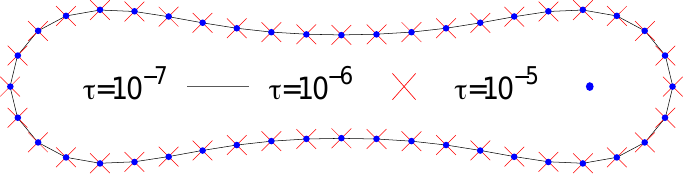}\label{fig:evolutionOfShapes-d}}
\subfigure[t=0.001]{\includegraphics[width=0.3\textwidth]{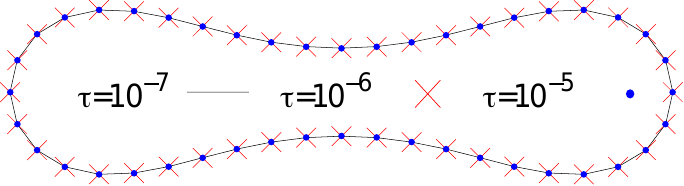}\label{fig:evolutionOfShapes-e}}
\subfigure[t=0.012]{\includegraphics[width=0.3\textwidth]{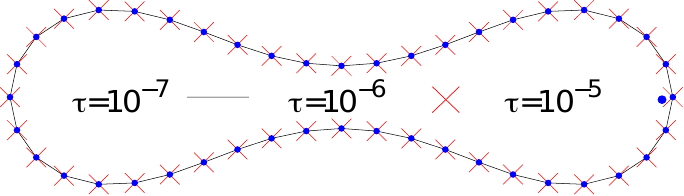}\label{fig:evolutionOfShapes-f}}

\subfigure[t=0.0002]{\includegraphics[width=0.3\textwidth]{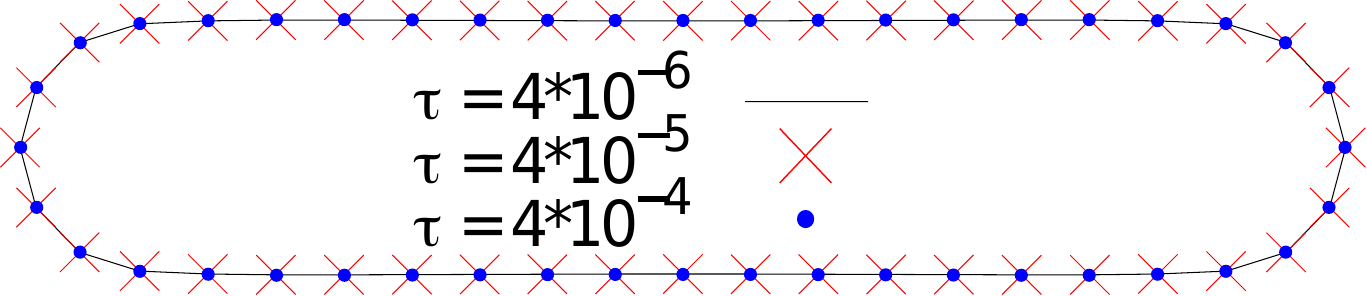}\label{fig:evolutionOfShapes-g}}
\subfigure[t=0.0008]{\includegraphics[width=0.3\textwidth]{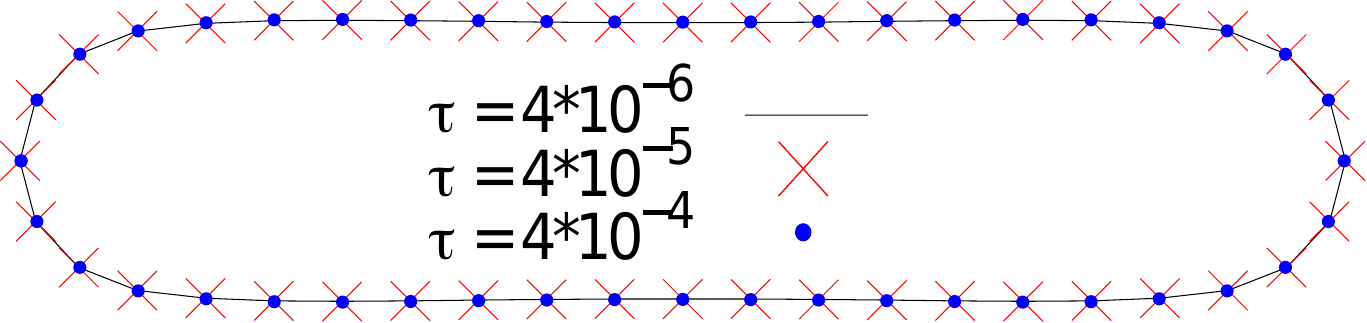}\label{fig:evolutionOfShapes-h}}
\subfigure[t=0.01]{\includegraphics[width=0.3\textwidth]{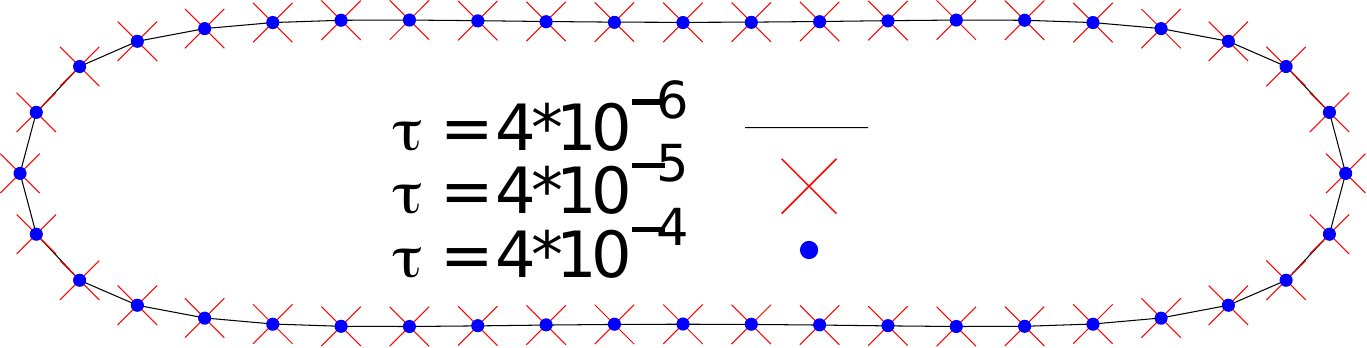}\label{fig:evolutionOfShapes-i}}

\subfigure[]{\includegraphics[width=0.33\textwidth]{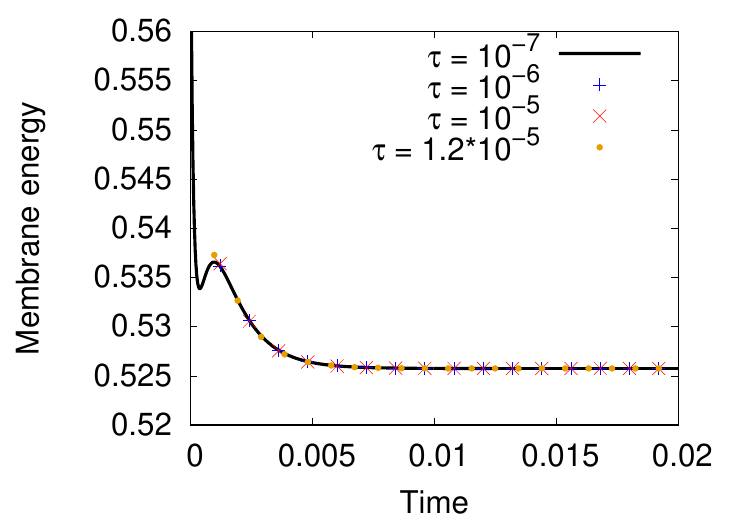}\label{fig:evolutionOfShapes-j}}
\subfigure[]{\includegraphics[width=0.33\textwidth]{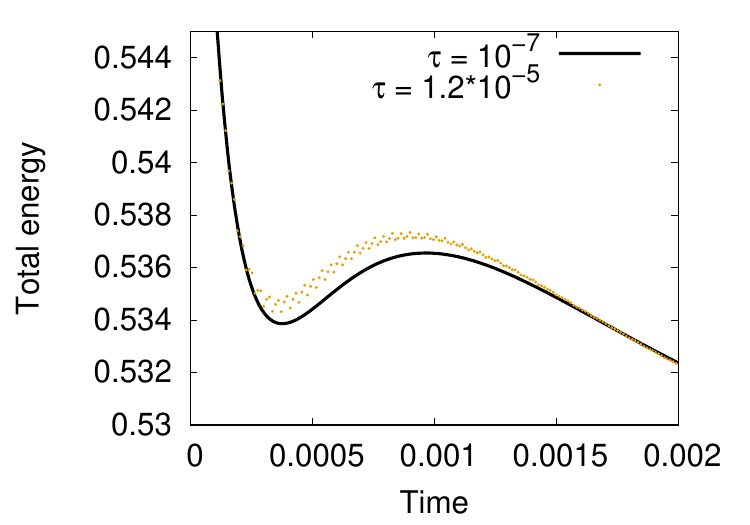}\label{fig:evolutionOfShapes-k}}
\label{fig:evolutionOfShapes}
\caption[]{Evolution of the shell cross section shapes for (a)-(c) the surface tension stiffness dominant case, and (d)-(f) the bending stiffness dominant case, and (g)-(i) the stretching dominant case with different time step sizes. 
Membrane energy over time is shown for the bending stiffness dominant case (j), with a close-up on the local peak (k).}
\end{figure}

According to the previous section, it seems sufficient to do further studies using the $h_1$-mesh. The three different cases are now being analyzed using different time step sizes. \autoref{fig:evolutionOfShapes}(a-c) illustrates the shape evolution of the surface tension case (tested with timesteps $\tau=2.5\cdot 10^{-4}, 2.5\cdot 10^{-5}, 2.5\cdot 10^{-6}$).
During the evolution, slight differences in the shapes can be observed. The stationary state shows very good agreement with the expected spherical shape (\autoref{fig:evolutionOfShapes-c}), with only slight tangential differences (surface tension works in normal direction only), for larger time steps. 
The simulation with the largest time step required only $80$ time steps to reach the stationary state. The necessary compute time of less than 1 minute on a single core CPU (Intel Haswell, $2.50\,$GHz) illustrates the efficiency of the proposed method. 

In the bending stiffness dominant case (case 2) the shape differences are also comparatively small. Time step sizes $10^{-5}, 10^{-6}, 10^{-7}$ have been tested. \autoref{fig:evolutionOfShapes}(d-f) shows the time evolution of the cross section shape. The agreement of point coordinates is good enough to have no visible difference between the different time step sizes. This is also observable when analyzing the membrane energy (\autoref{fig:evolutionOfShapes}(j,k)) 
\begin{align}
E_{\text{membrane}} = 2\pi\int_{\Gamma} R\left(E_{\text{tension}}+E_{\text{bend}}+E_{\text{stretch}}\right) \text{d}A\, .
\end{align}
\autoref{fig:evolutionOfShapes-j} shows $E_{\text{membrane}}$ for the bending stiffness dominant case for all tested time step sizes. A close up view of the increase at $t=0.001$ is shown in \autoref{fig:evolutionOfShapes-k}, for the largest and smallest time step. Additionally, the time step $\tau=1.2\cdot 10^{-5}$ is included. This value was the largest time step size, where stable simulations were possible, as the explicit coupling between  flow and membrane elasticity introduces a numerical stiffness. The membrane energy in the case of $\tau=1.2\cdot 10^{-5}$ oscillates for  $t<0.0015$. These oscillations smooth out when the energy dissipates, leading to the exact same behaviour as for the smallest time step.
Consequently, even with large time steps, the membrane energy dissipates in the same manner as with small time steps and the resulting shapes show nearly no differences. However, the coupling of bending stiffness and surface elasticity makes the membrane movement more subtle, such that it takes relatively long to reach the stationary state. With the largest time step, $2000$ time steps were required. The total compute time in this case was $\approx 30$ minutes. 
 
The stretching dominant case has been tested with time steps $\tau = 4\cdot 10^{-5}, 4\cdot 10^{-6},4\cdot 10^{-7}$. Images are shown in \autoref{fig:evolutionOfShapes}(g-i). As for the bending stiffness dominant case, the membrane energy dissipates in the same manner for small and large time steps and there are only very small shape differences between the shapes for the respective time step sizes. The simulation required $250$ time steps to reach the stationary state. The total compute time in this case was $\approx 3$ minutes on a single core CPU.

A convergence study is done also for the time step analysis. The results can be seen in \autoref{tab:EOC} and show the expected first order convergence. 

\begin{table}
\begin{tabularx}{\textwidth}{X X X X | X X X X}
&\multicolumn{3}{c}{mesh study} &&\multicolumn{3}{c}{time step study}  \\
Case & 1 & 2 & 3 & &1 & 2 & 3 \\
\hline\hline
$E^{h_1}$ & $2.64\text{E}-3$ & $1.39\text{E}-3$
 & $3.94\text{E}-4$ & $E^{\tau_1}$ & $2.77\text{E}-3$ & $6.56\text{E}-7$
 & $2.06\text{E}-3$\\
 $E^{h_2}$ & $1.32\text{E}-3$ & $6.36\text{E}-4$
 & $1.67\text{E}-4$ & $E^{\tau_2}$& $2.89\text{E}-4$ & $8.53\text{E}-8$
 & $1.86\text{E}-4$\\
 \hline
$E_P^{h_1}$ & $4.29\text{E}-2$ & $4.90\text{E}-3$
 & $4.90\text{E}-3$ & $E_P^{\tau_1}$& $1.45\text{E}-4$ & $7.04\text{E}-4$
 & $2.16\text{E}-5$\\
 $E_P^{h_2}$ & $1.70\text{E}-3$ & $1.00\text{E}-3$
 & $1.25\text{E}-3$ & $E_P^{\tau_2}$& $1.77\text{E}-5$ & $7.00\text{E}-5$
 & $2.37\text{E}-6$\\
  \hline\hline
 \textbf{EOC}$_{\textbf{E}}$ & 0.99 & 1.13 & 1.24 & \textbf{EOC}$_{\textbf{E}}$ & 0.98 & 0.89 & 1.04 
\\ \hline
 \textbf{EOC}$_{\textbf{P}}$ & 1.96 & 2.27 & 1.98  & \textbf{EOC}$_{\textbf{P}}$ & 0.92 & 1.00 & 1.01 
\end{tabularx}
\label{tab:EOC}
\caption{EOC for the different test cases}
\end{table}

\subsection{Timestep stabilization of the stretching force}
In some applications, the time step restrictions due to the surface forces can be rather strict.
In general, such restrictions appear when the elastic force is stiff, and the forces are explicitly calculated from the previous membrane position. To overcome this problem, implicit and semi-implicit schemes have been presented to compute the surface tension and bending forces \cite{Lai2019}. 
The idea is to include the interface movement monolithically in the flow solver to evaluate the surface forces implicitly at the newly computed time step.
This approach has been shown to work extremely well in the immersed boundary method, and can in principle be combined with our present ALE method.

However, while this methodology has been proposed for surface tension and bending forces, we are not aware of a similar method for the in-plane elastic stretching force. Accordingly, we present a novel approach to monolithically couple the in-plane stretching force to the flow in this section. 

The crucial issue in the stretching force (\autoref{eq:stretchingForce}) is the implicit treatment of the principal stretches $\lambda_1$ and $\lambda_2$. A prediction for the values of the principal stretches in the upcoming time step can be obtained from the following evolution equations \cite[Appendix]{Mokbel}:
\begin{align}
\partial_t^\bullet\lambda_1 - \lambda_1\nabla_{\Gamma}\cdot\mathbf{v} = 0,\quad   \partial_t^\bullet\lambda_2 - \lambda_2 \frac{v_r}{R} = 0\,  \quad \text{on }\Gamma,
\label{eq:principalStretchesImplicit1}
\end{align}
where $\nabla_\Gamma$ is the (non-axisymmetric) surface divergence in the two-dimensional domain. 
Accordingly, we can approximate the new values of the principle stretches by 
\begin{align}
\lambda_1^{n} = \lambda_1^{n-1} + \tau \lambda_1^{n-1}\nabla_{\Gamma}\cdot\mathbf{v}^{n},\quad   
\lambda_2^n = \lambda_2^{n-1} +\tau \lambda_2^{n-1} \frac{v_r^n}{R} = 0\,  \quad \text{on }\Gamma
\label{eq:principalStretchesImplicit2}
\end{align} 
where the old solution $\lambda_i^{n-1}$ is calculated, as before, from the current point coordinates (see \autoref{eq:principalStretchesDiscrete}) on $\Gamma$.

These equations can now be solved together with the Navier-Stokes equation in one system. 
To complete the monolithic coupling, the term $\left(\frac{\partial E_{\text{stretch}}}{\partial\Gamma}\right)^{n-1}$ in \eqref{eq:weakForm} is replaced by 
\begin{align}
\left(\frac{E_{\text{stretch}}}{\partial\Gamma}\right)^n &:= \left(\kappa^{n-1}\,\textbf{n}^{n-1} -\nabla_{\Gamma}\right)\left[\left(K_A+K_S\right)\left(\lambda_1^n - 1\right) + \left(K_A-K_S\right)\left(\lambda_2^n - 1\right)\right] - \frac{2K_S}{R}\left(\lambda_1^n-\lambda_2^n\right)\begin{pmatrix} 0 \\ 1 \end{pmatrix}\, , \label{eq:stretchingForceImplicit}
\end{align}
which includes the implicitly calculated values of the principal stretches.

\begin{figure}
	\centering 		
	\subfigure[$h_1$]{\includegraphics[width=0.32\textwidth]{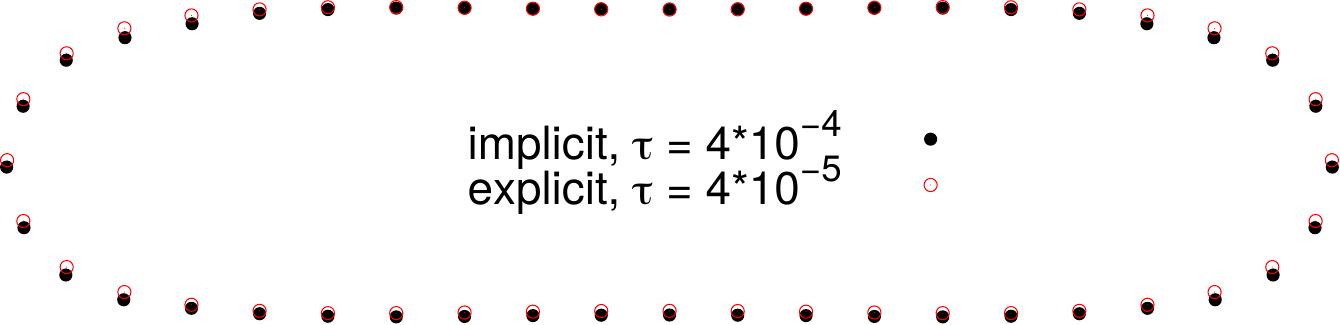}\label{fig:CompareExVsIm-a}}		
	\subfigure[$h_2$]{\includegraphics[width=0.32\textwidth]{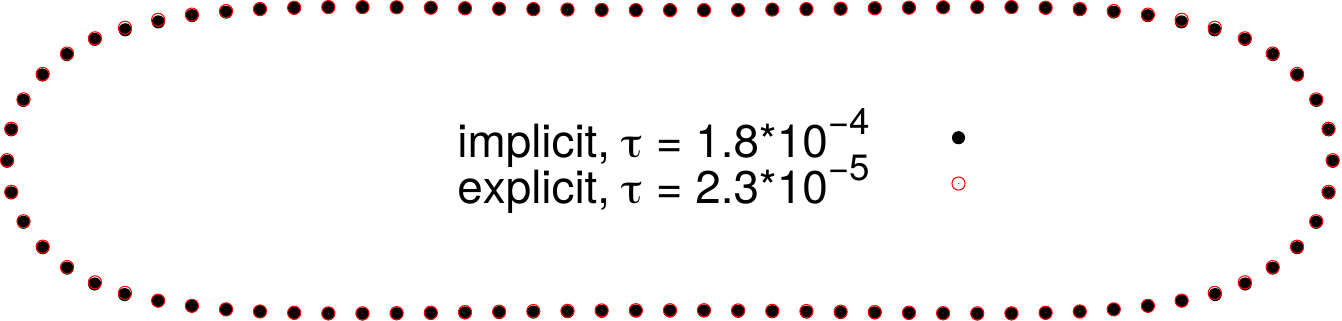}\label{fig:CompareExVsIm-b}}	
	\subfigure[$h_3$]{\includegraphics[width=0.32\textwidth]{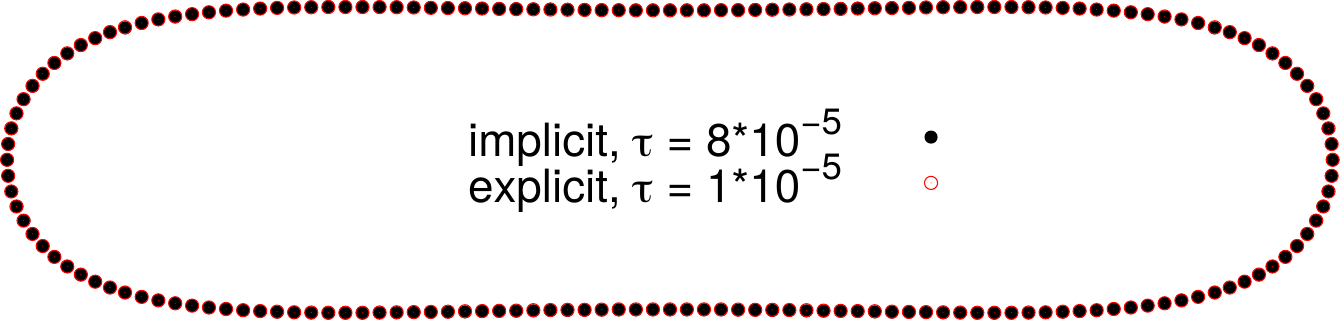}\label{fig:CompareExVsIm-c}}
	\label{fig:CompareExVsIm}
	\caption[]{Shapes of explicit vs. implicit simulations at $t=0.01$ for the different mesh resolutions. The shape difference vanishes for finer meshes. A time step enlargement of factor 8-10 is possible when using the implicit stretching force approach.}
\end{figure}

The capability of this implicit strategy has been tested with the parameters of the stretching dominant case. \autoref{fig:CompareExVsIm} shows the grid points of the membrane for the three different meshes with sizes $h_1$, $h_2$, and $h_3$ and different time step sizes. The time step sizes have been chosen to be maximal in each case, i.e. such that the usage of a larger time step would lead to crashing simulations due to oscillations on the membrane points.  

As seen in \autoref{fig:CompareExVsIm} the implicit approach permits an enlargement of the time step size by a factor of $8-10$. 
Also notable is that the maximum time step is roughly proportional to the grid size, for both, implicit and explicit approach. 
Only for the coarsest mesh (\autoref{fig:CompareExVsIm-a}), differences between the explicit and implicit approach are visible, which can be attributed to the large time step sizes. 
Hence, in particular for higher mesh resolution at the membrane, the implicit approach can lead to high quality results with up to 10 times faster simulations.

\subsection{Uniaxial compression of a biological cell}

As mentioned in the Introduction \autoref{sec:Introduction}, the ALE method presented in this work offers the opportunity to restrict the simulation to only one of the fluid phases. This capability, which is not present in many other methods (e.g. Immersed Boundary Methods), can be very useful if the viscosity ratio between the two fluid phases is large such that the phase with the smaller viscosity has little influence on the results and can be neglected. This does not only speed up the simulations but can also greatly simplify the coupling of the elastic surface to exterior forces, as for example in a contact problem. 

We illustrate this in the following, as we consider a fluid-filled elastic shell under external compression. The application behind this test case is the uniaxial compression of a biological cell during an atomic force microscopy (AFM) experiment, which is used to measure cell mechanical properties. The viscosity of the intracellular fluid is typically several orders of magnitude larger than the viscosity of the surrounding medium. A detailed description of the problem can be found in our work \cite{poisson2019}.

\begin{figure}
	\centering 		
	\subfigure[]{\includegraphics[width=0.35\textwidth]{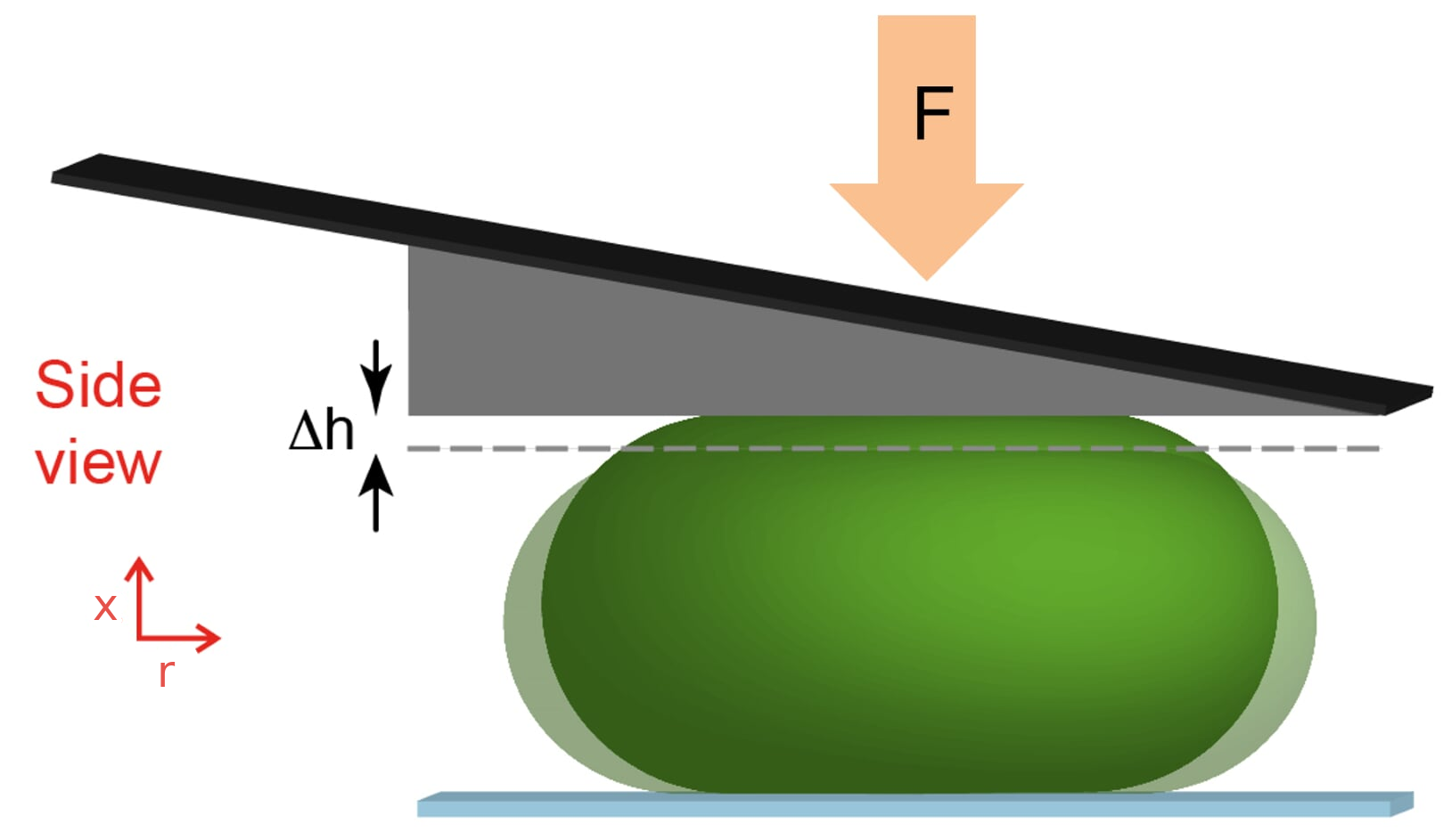}\label{fig:AFMCantilever-a}}
	\subfigure[]{\includegraphics[width=0.28\textwidth]{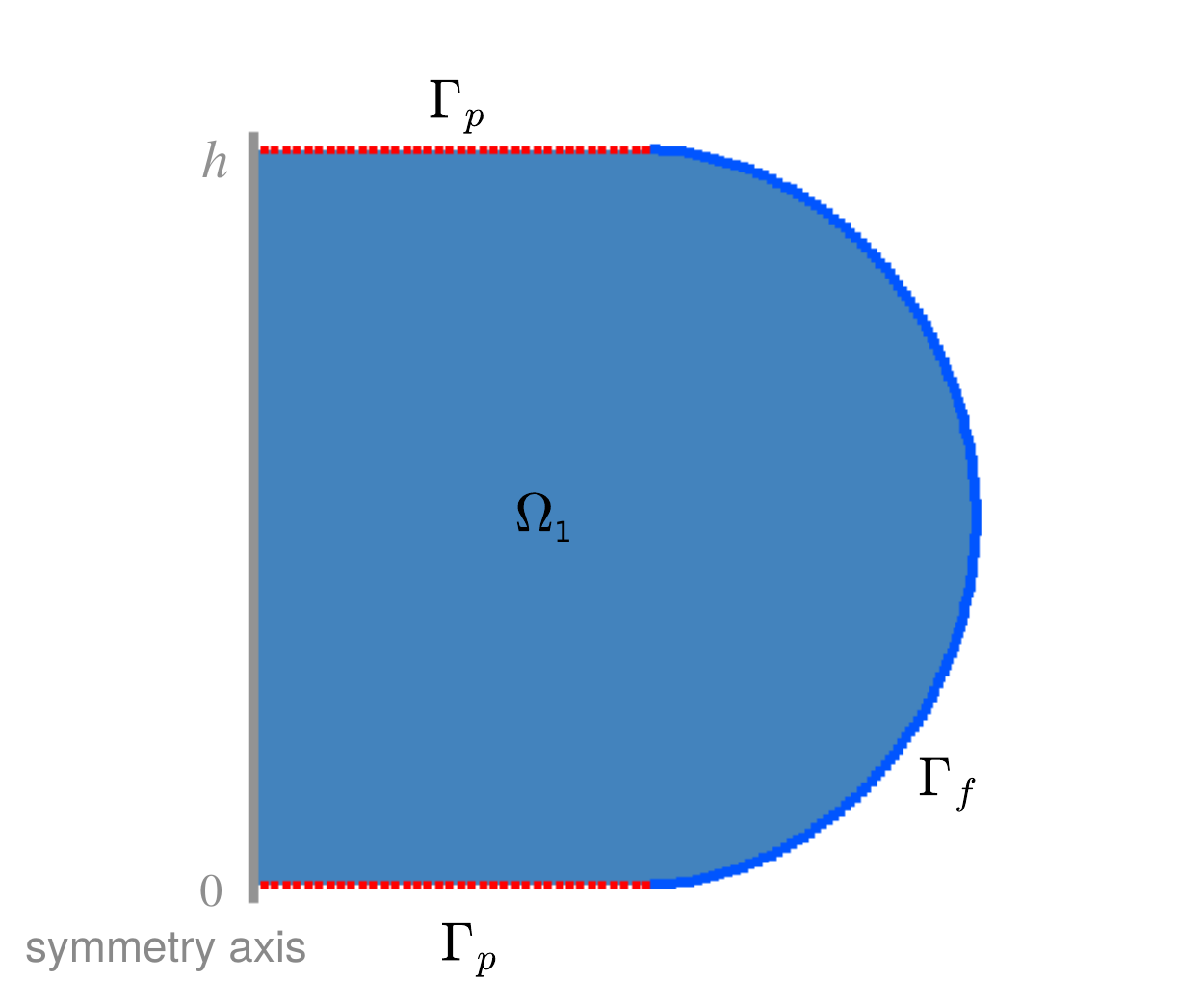}\label{fig:AFMCantilever-b}}
	\subfigure[]{\includegraphics[width=0.3\textwidth]{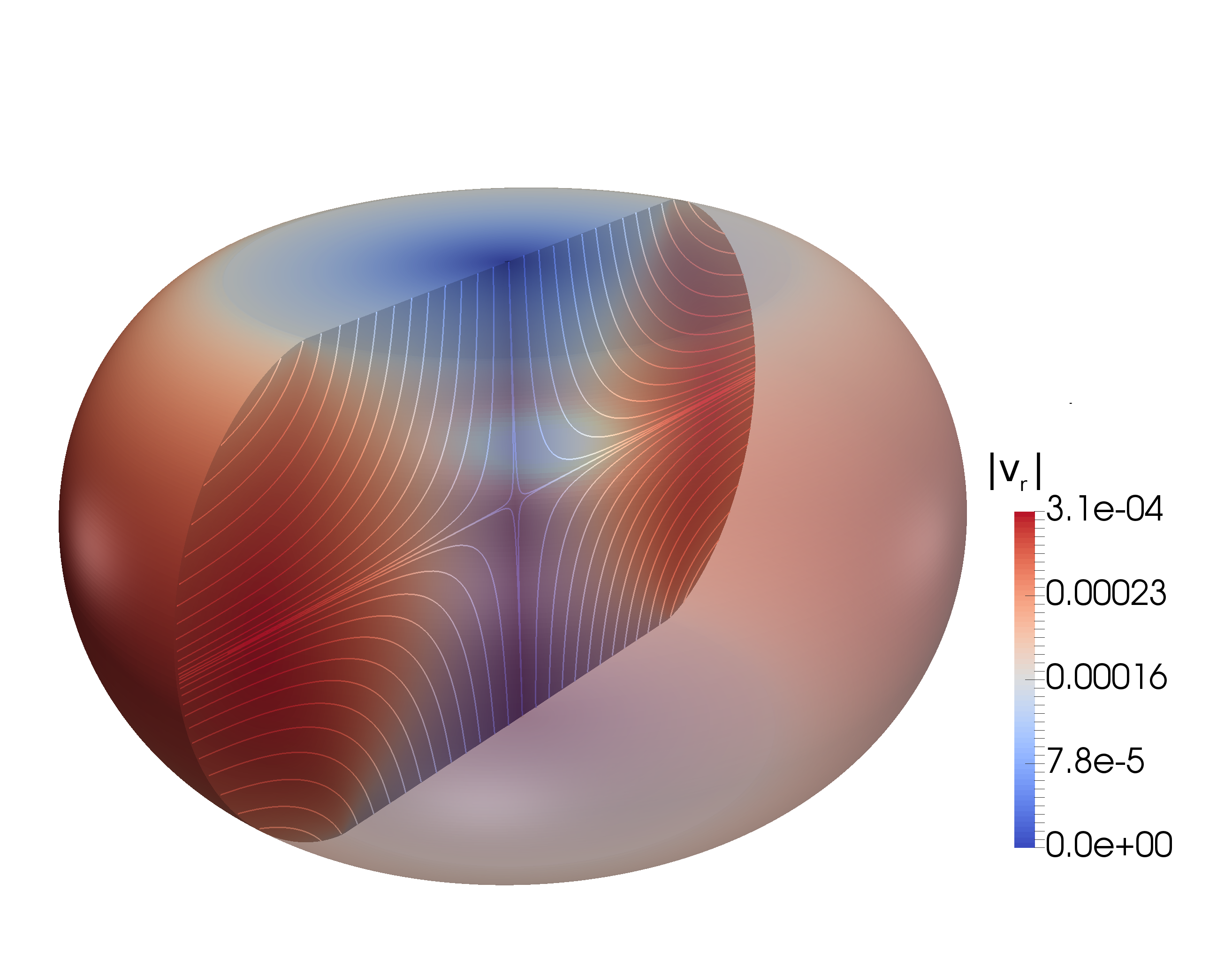}\label{fig:AFMCantilever-c}}
	
	\subfigure[]{\includegraphics[width=0.3\textwidth]{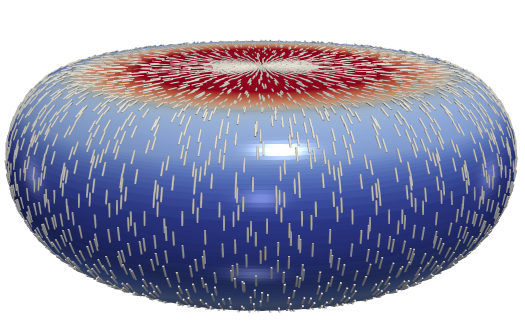}\label{fig:AFMCantilever-d}}
\subfigure[]{\includegraphics[width=0.3\textwidth]{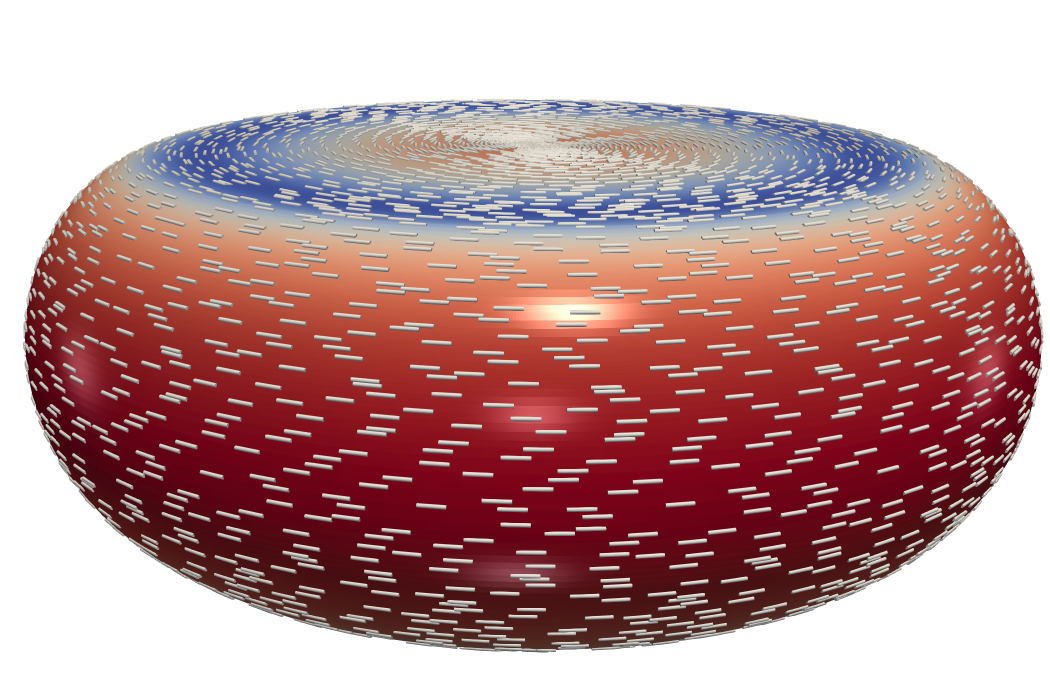}\label{fig:AFMCantilever-e}}
	
	\label{fig:AFMCantilever}
	\caption[]{(a) Experimental setup of a biological cell under uniaxial compression. (b) The simulation domain. The plates here are on the top and bottom sides of the cell with the boundary $\Gamma_p$ (red dotted line). The free part of the cell boundary is $\Gamma_f$ (blue line). (c) Streamlines during the compression process, colored by $|v_r|$. (d,e) Compressed cell colored by the values of the principal stretches $\lambda_1$ (d) and $\lambda_2$ (e) together with tangential lines in the corresponding stretch directions. }
\end{figure}

Consider a rounded biological cell confined between two parallel plates. During the experiment, the upper plate moves with a prescribed velocity $v_{\rm compress}$ until the initial plate distance $h(t)=h_0$ is decreased over time to $h(t)=h_0-\Delta h$, see \autoref{fig:AFMCantilever}(a). The necessary force $F$ is constantly measured during compression. Matching experimental data with simulations can be used to extract the surface properties (i.e. $K_A$ and $K_S$) of the cell's elastic shell (the actin cortex). The experimental setup along with the simulation domain is illustrated in \autoref{fig:AFMCantilever}(a,b).

In the simulations, the interior of the cell is denoted by  $\Omega_1$ which is bounded by the elastic cell cortex $\Gamma$ and the symmetry axis. $\Gamma$ is subdivided into the area touching the plates $\Gamma_p$ and the free surface area $\Gamma_f$. 
During compression a part of the free surface will touch the plate, accordingly $\Gamma_p$ and $\Gamma_f$ are time-dependent:
\begin{align}
\Gamma_p(t) = \left\lbrace \mathbf{x}=(x,r)\in\Gamma: x=0 \lor x=h(t) \right\rbrace, \qquad \Gamma_f(t) = \Gamma/\Gamma_p(t).
\end{align} 
A simple contact algorithm is implemented in the numerical simulation: Surface grid points are initially marked to belong to either $\Gamma_p$ or $\Gamma_f$, as soon as a grid point of $\Gamma_f$ touches the upper plate ($x\geq h(t)$), it is shifted to $\Gamma_p$.
The interface curve of $\Gamma$ for the initial meshes with plate distance $h=h_0$ is given by a minimal surface calculated according to equations described in \cite{FischerFriedrich2014}. 

The surrounding medium is neglected to avoid the complicated numerical handling of it being squeezed out of the contact region. Accordingly, the Navier-Stokes equations \eqref{eq:NavierStokes}-\eqref{eq:incompressibility} are only solved in $\Omega_1$. The corresponding  boundary conditions \autoref{eq:boundaryConditions} have to be adapted 
\begin{align}
\left(-p_1\mathbf{I} + \eta_1\left(\nabla\mathbf{v} + \left(\nabla\mathbf{v}\right)^T\right)\right)\cdot\mathbf{n} &=  \frac{\partial E}{\partial\Gamma} & \text{on}\ \Gamma_f\\
\mathbf{t}\cdot\left(-p_1\mathbf{I} + \eta_1\left(\nabla\mathbf{v} + \left(\nabla\mathbf{v}\right)^T\right)\right)\cdot\mathbf{n} &= \mathbf{t}\cdot \frac{\partial E}{\partial\Gamma}, & \text{on}\ \Gamma_p\\
v_x &= \delta_{x>0}\, v_{\rm compress} , & \text{on}\ \Gamma_p,
\end{align}
where ${\bf t}$ is the tangential vector to $\Gamma_p$.
 
Numerical studies have been conducted in the realistic physical parameter regime $h_0\in[6\, \mu$m$,16\,\mu$m$],\ \gamma\in[0.5\, $mN/m$,3\, $mN/m$],\ K_A=25\,$nN/$\mu$m$,\ K_S\in[8\,$nN/$\mu$m$,25\,$nN/$\mu$m$],\ K_B\in[0.11\,$nN/$\mu$m$,0.17\,$nN/$\mu$m$],\ \eta_1=1$Pa$\cdot$s. 
\autoref{fig:AFMCantilever-c} shows exemplary streamlines during compression, as fluid is driven towards the free boundary $\Gamma_f$, extending the cell's radius.
Detailed numerical results can be found in \cite{poisson2019}.

As shown in \cite{poisson2019} the simulations reproduce the characteristic force response of biological cells. Matching simulations with experiments can provide new insights in cell mechanical properties. To this end, an extremely fine grid resolution with approximately 2000 grid points along the surface contour is necessary to disentangle contributions from surface tension and surface elasticity (see \cite{poisson2019}). This fine grid resolution makes the problem unfeasible to full 3D simulations and underlines the relevance of the proposed method. 

\subsection{Shape oscillations of novel microswimming shells}

Finally, we illustrate the versatility of our method by simulating hollow elastic shells immersed in a fluid, i.e. a combination of a compressible fluid inside and an incompressible fluid outside an elastic shell. 
Such shells have been recently proposed in \cite{Djellouli2017} as a very powerful new mechanism for microscopic swimming, since they are extremely fast, simple in shape and controllable from the outside (e.g. by ultrasound-driven pressure oscillations).

\begin{figure}
	\centering 		
	\subfigure[]{\includegraphics[height=0.3\textwidth]{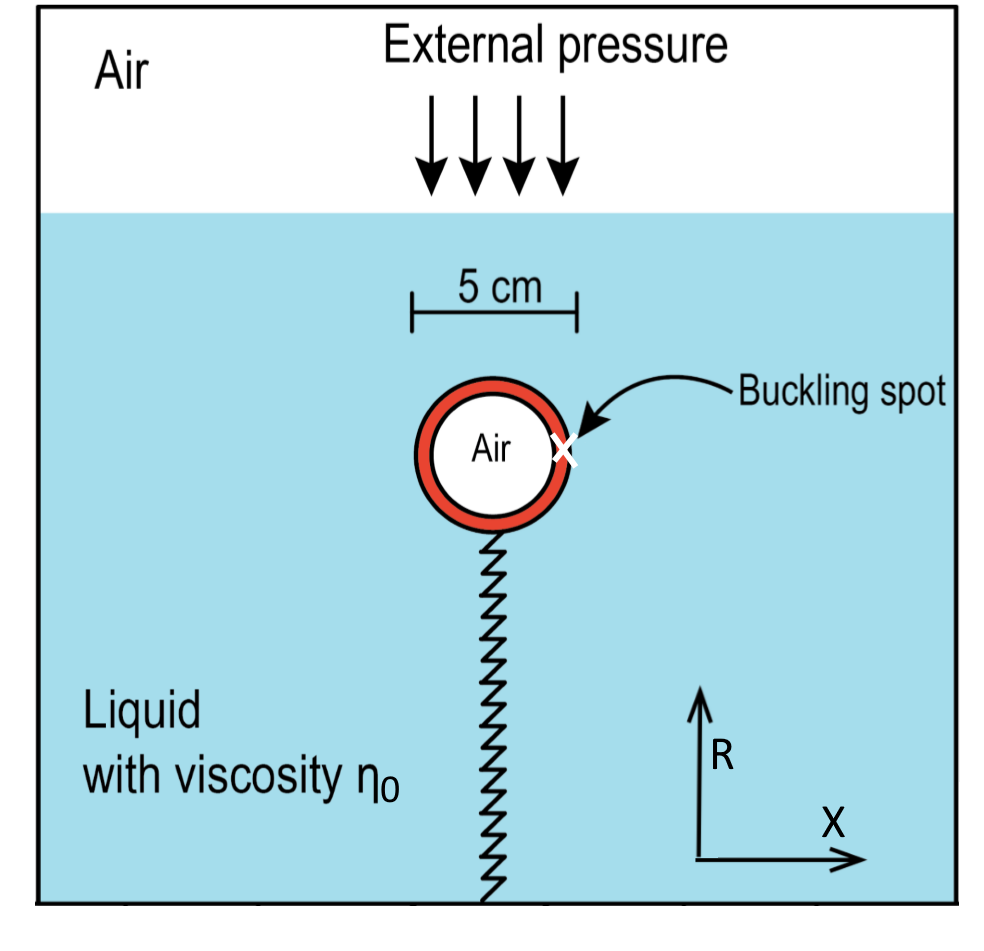}\label{fig:microswimmer-a}}\hspace{15px}
	\subfigure[]{\includegraphics[height=0.3\textwidth]{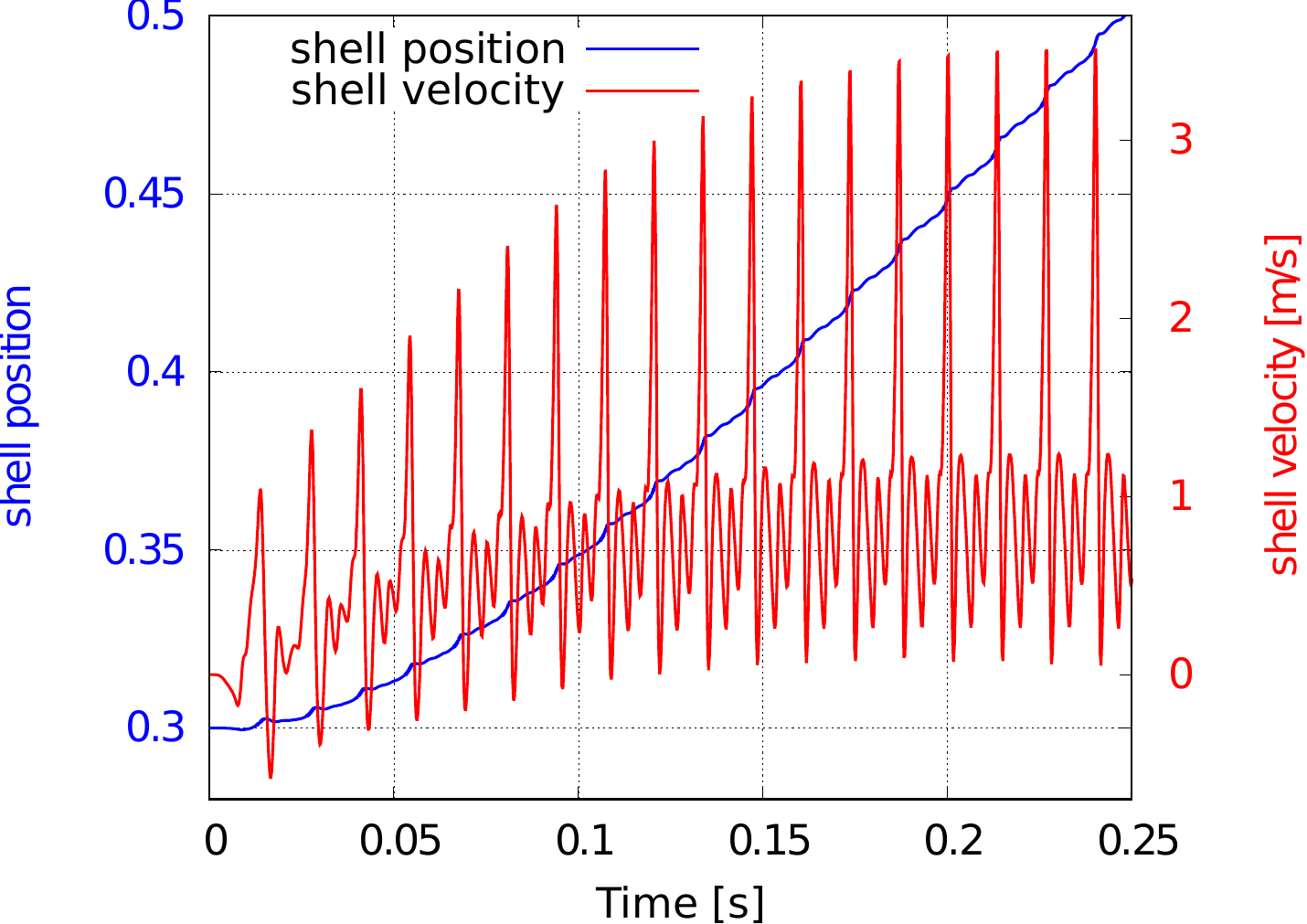}\label{fig:microswimmer-graph}}
	
	\subfigure[$t=0.003\,$s]{
	\includegraphics[width=0.2\textwidth]{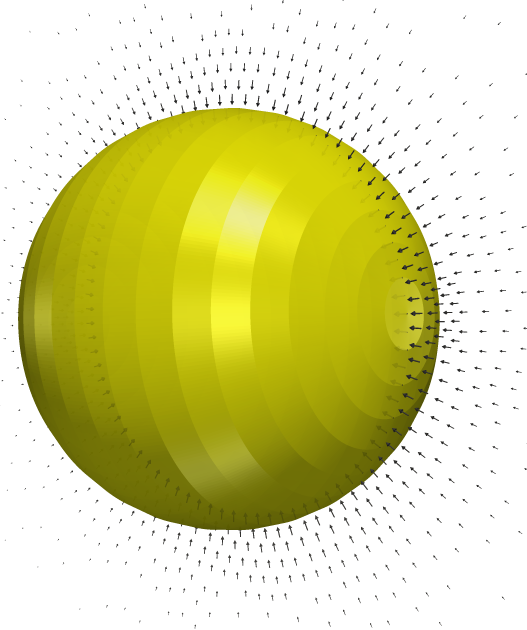}
	\label{fig:microswimmer-1}}
		\subfigure[$t=0.008\,$s]{\includegraphics[width=0.21\textwidth]{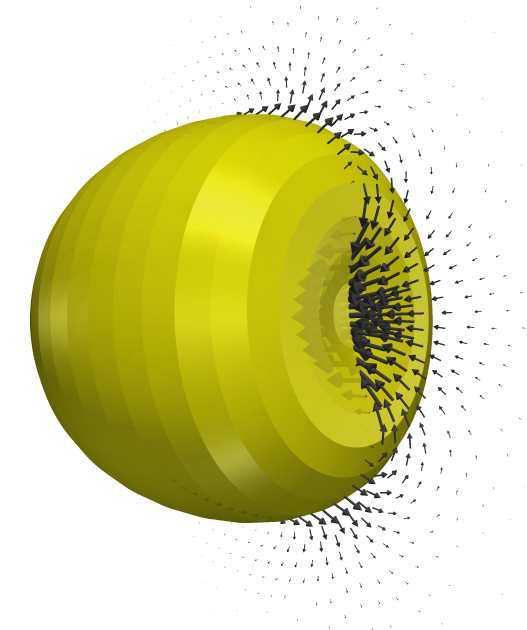}\label{fig:microswimmer-2}}
		\subfigure[$t=0.01\,$s]{\includegraphics[width=0.2\textwidth]{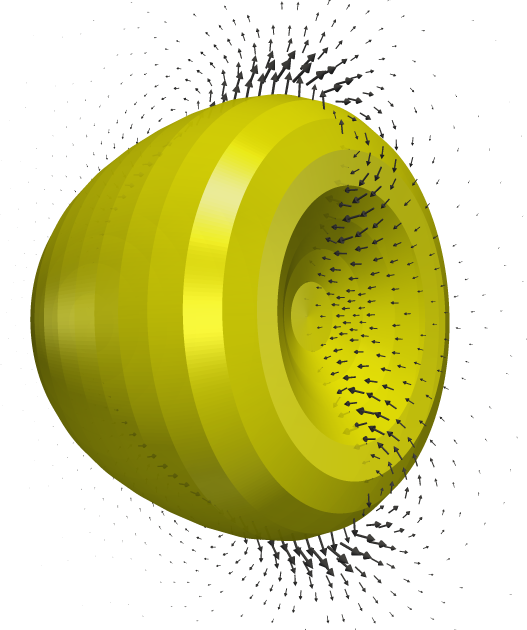}\label{fig:microswimmer-3}}
		\subfigure[$t=0.015\,$s]{\includegraphics[width=0.23\textwidth]{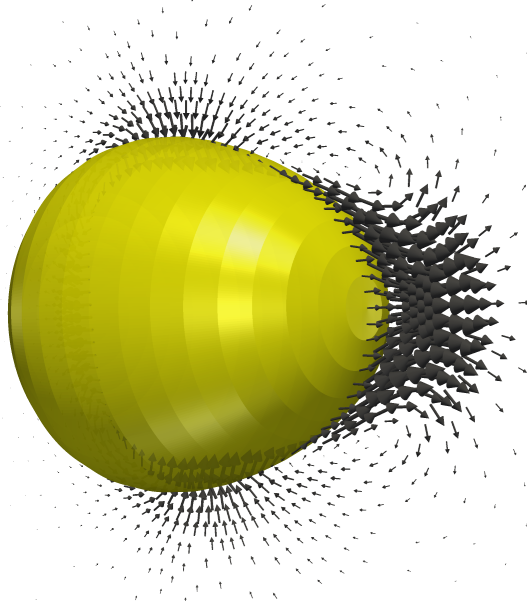}\label{fig:microswimmer-4}}
	\label{fig:Microswimmer}
	\caption[]{(a) Experimental setup of the shell under pressure oscillations, adapted from \cite{Adel_dissertation}. (b) Position and velocity of the shell's center of mass in $x-$direction during the first $0.25\,$s of acceleration. The frequency of the outer pressure oscillations is $75\,$Hz. (c)-(f) Snapshots of the buckling shape dynamics during one cycle in the simulations, together with the velocity field. }
\end{figure}

The idea of the propulsion mechanism is as follows. 
Consider a microscopic spherical elastic shell, filled with air and immersed in an viscous liquid. The shell has a weak spot, i.e. a small area, where its (otherwise uniform) thickness  is slightly reduced. The outer fluid pressure is assumed to be controllable, e.g. via ultrasound, see Fig.~\ref{fig:microswimmer-a}. 
If the outer pressure is increased, the air inside the shell will be compressed such that the shell shrinks but remains spherical. When the pressure difference exceeds a shell-specific threshold, the shell will buckle at the weak spot \cite{Djellouli2017}, which is a very rapid process accompanied by rapid elastic surface oscillations. Decreasing the outer pressure again leads to inflation of the shell and debuckling. The motion of the shell during such a buckling cycle induces a fluid flow moving the  shell in direction of the weak spot. 

Here, we present the first numerical simulations of this process, which can help to gain a better understanding of the buckling dynamics and the influence of parameters on the swimming efficiency. 
The sudden release of stored elastic energy during the buckling of the shell leads to very high flow rates such that the hydrodynamics are significantly influenced by inertial forces. Hence, even at the microscale, the full \textit{Navier}-Stokes equations are necessary to describe the process \cite{Djellouli2017}.
The simulation of the inner phase $\Omega_1$ is not necessary due to the low density and viscosity of air. Instead, we assume a homogeneous air pressure $p_1$ inside and use adiabatic gas theory, to relate this pressure to the inner shell volume $V$ by $p_1 = p_{1,0}\,(V_0/V)^{1.4}$, where $p_{1,0}$ and $V_0$ denote the respective initial values. 
Accordingly, the stress exerted by the air is reduced to $p_1 {\bf n}$ whereupon the stress boundary condition \eqref{eq:boundaryConditions} becomes
\begin{align}
\left(-p_0\mathbf{I} + \eta_0\left(\nabla\mathbf{v} + \left(\nabla\mathbf{v}\right)^T\right)\right)\cdot\mathbf{n} &=  -p_{1,0}\,(V_0/V)^{1.4} {\bf n} - \frac{\partial E}{\partial\Gamma}.
\end{align}

The weak spot is introduced by slightly decreasing the elastic surface moduli locally around the membrane point touching the symmetry axis ($R=0$) on the right. Note, that the numerical results are independent of the exact amount of this reduction, as the weak spot only serves as a locator for the occurring buckling instability.
Numerical parameters are as in \cite{Djellouli2017} where a macroscopic shell diameter of $\approx 5\,$cm and increased viscosity (glycerol $\eta = 1\,$Pa or oil $\eta = 37.5\,$Pa) was used to mimic flow conditions at the microscale. The initial air pressure is $p_{1,0}=1\,$bar.

During the simulation, an external pressure difference is imposed using a sinusoidal function (oscillating between $p_0=1.0\,$bar and $p_0=1.77\,$bar) at the outer boundaries of the computational domain. 
The change in outer pressure leads to periodic buckling and debuckling of the microswimmer shell. 
Images of different stages of a buckling cycle are shown in \autoref{fig:microswimmer-1}-(f) together with the surrounding fluid velocities. 
Increasing pressure in the beginning leads to uniform shrinkage (\autoref{fig:microswimmer-1}) and buckling (\autoref{fig:microswimmer-2}-(e)). The concomitant complex flow patterns induce a thrust of the shell, propelling it to the right. Decreasing pressure leads to inflation and debuckling and a further propulsion to the right, see \autoref{fig:microswimmer-4}. Shape evolution and flow patterns are comparable to the experiments in \cite{Djellouli2017}.

The rich dynamics during this swimming process require a great number of time steps to be resolved. This is illustrated in Fig.~\ref{fig:microswimmer-graph} which shows the position and velocity of the microswimmer shell during the first 0.25 seconds. Fast elastic surface oscillations occur due to the high eigenfrequency of the swimmer (see Fig.~\ref{fig:microswimmer-graph}, red curve). To resolve these oscillations which are accompanied by complex flow patterns we use a time steps size of $\tau=2.5\,\mu$s.
Accordingly, $10^5$ time steps are needed to accurately resolve the acceleration process for the given parameters. For other excitation frequencies and amplitudes, the acceleration process may even take significantly longer as a more complex interplay of excitation frequency and eigenfrequency develops. The corresponding number of necessary time steps makes full 3D simulations illusive.
Axisymmetric immersed boundary methods cannot cope with different physical processes inside and outside the shell (compressible/incompressible fluid) and axisymmetric boundary element methods are restricted to the Stokes regime.
Hence the method proposed here is, to our knowledge, the only available method to simulate the microscopic swimming in reasonable times.

Our simulations shall be used in the future in collaboration with the authors of \cite{Djellouli2017} to gain understanding of the complex dynamical coupling of surface shape deformations, pressure oscillations and shell propulsion to create efficient novel microswimmers.

\section{Conclusion}
In this paper we have presented a novel ALE method to simulate axisymmetric elastic surfaces immersed in Navier-Stokes fluids. 
As inherent to ALE methods, the grid is matched to the elastic material which can therefore be resolved with relatively few grid points. The axisymmetric setting reduces the system effectively to a two-dimensional problem. Elastic surface forces are discretized with surface finite-differences and coupled to evolving finite elements of the bulk problems. An implicit coupling strategy reduces time step restrictions induced by the stiffness of the stretching elasticity.

The method combines high accuracy with computational efficiency, which is confirmed in several numerical test cases dominated either by surface tension, bending stiffness or in-plane elasticity. In all these cases we find that numerical errors are relatively small even for relatively coarse grids and converge with order 1-2 with respect to grid size and time step size. The computational times of the test problems are on the order of minutes on a single core CPU. 
While such a high computational speed is not required in most typical applications, it does enable otherwise unfeasible simulations as soon as the problems involve a high number of grid points or rich dynamics demanding many time steps. 

As examples, we present first simulations of the observed shape oscillations of novel microswimming shells \cite{Djellouli2017} and the uniaxial compression of biological cells filled with cytoplasm \cite{poisson2019}. 
In collaboration with (bio)physicists, the method is currently used to get more insight into the dynamics of microswimming shells and the cellular cortex. 
As a next step we plan to extend the method to full three dimensions. 

\section*{Acknowledgement}
We acknowledge support from the German Research Foundation DFG (grants AL1705/3, AL1705/5) and the DFG Research Unit FOR 3013. 
Simulations were performed at the Center for Information Services and High Performance Computing (ZIH) at TU Dresden.

\newpage
\section{Appendix}\label{appendix}
\subsection{Implementation details I - mesh and boundary conditions}
The presented method is implemented in the finite element toolbox AMDiS \cite{Vey2006, Witkowski2015}. 
While AMDiS does support Taylor-Hood (P2/P1) elements, it does not support different solution variables being defined on different meshes, as necessary for the pressure $p_i$ here. Accordingly, we use some technical workarounds to implement the proposed method, described in the following. 

\begin{figure}
	\centering 		
	\includegraphics[width=0.6\textwidth]{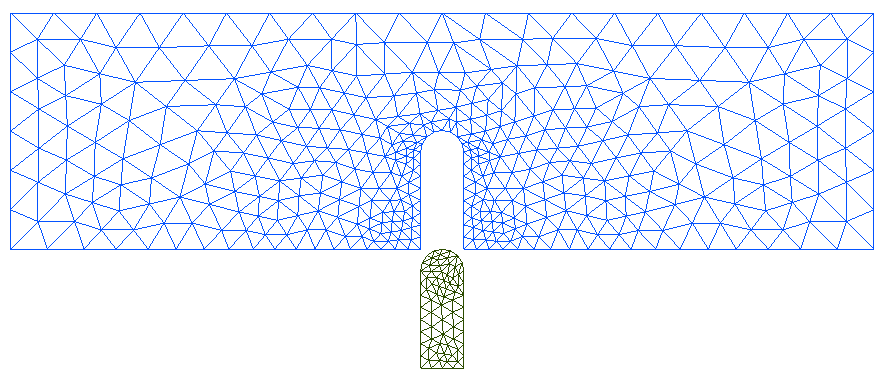}\label{fig:loadedMesh}
	\caption[]{Mesh as it is imported by the simulation.}
\end{figure}

A single mesh that contains both domains $\Omega_0, \Omega_1$ as seperate disconnected parts is constructed in gmsh \cite{gmsh}. 
To easily distinguish between internal and external fluid, the internal part is translated below the symmetry axis ($r<0$). \autoref{fig:loadedMesh} shows the mesh in the initial state.
At the beginning of the simulation an indicator function is created discriminating between the two mesh parts based on the $r-$value (grid points with negative $r$ belong to the internal fluid). 
Immediately afterwards, the internal fluid mesh is shifted upwards to obtain a spatially matched grid, which is yet unconnected at the interface. 

The Navier-Stokes equations are assembled on both mesh partitions separately using Lagrangian P2/P1 elements. The implementation of interfacial stress balance \autoref{eq:boundaryConditions} and velocity continuity are technically realized as follows. 
The discrete membrane $\Gamma_h = \Gamma_{h,0}\cup\Gamma_{h,1}$ is composed of the membrane boundaries of the mesh for the external and internal fluid, respectively.  
Due to the disconnected grid, we can only enforce one-sided stress conditions, which we prescribe as follows
\begin{align} \label{eq:bc_stress0}
\left(-p_0\mathbf{I} + \eta_0\left(\nabla\mathbf{v} + \left(\nabla\mathbf{v}\right)^T\right)\right)\cdot\mathbf{n} &=  -\frac{\partial E}{\partial\Gamma} & \text{on }\Gamma_{h,0} \\
-\left(-p_1\mathbf{I} + \eta_1\left(\nabla\mathbf{v} + \left(\nabla\mathbf{v}\right)^T\right)\right)\cdot\mathbf{n} &=  0 & \text{on }\Gamma_{h,1} \label{eq:bc_stress1}
\end{align}
Define $n$ as the number of ($P2-$)DOFs on the membrane. Let $j=0,\dots,n-1$, be an ordered (e.g. counterclockwise) numbering of the DOFs on $\Gamma_{h,0}$ and $\Gamma_{h,1}$, such that the numbers of both boundaries are equal. 

\begin{figure}
	\centering 		
	\includegraphics[width=0.65\textwidth]{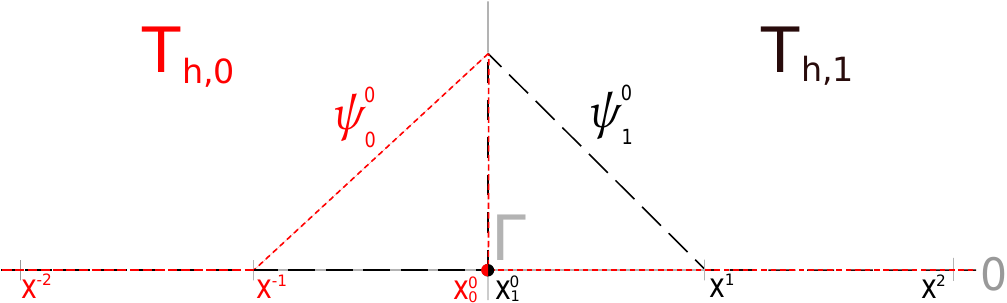}\label{fig:testFunctions}
	\caption[]{1D example for the test-functions across $\Gamma$. The grid coordinates are denoted with $\mathbf{x}^i$, where the membrane coordinate has the superscript $0$. Since it exists for both meshes, there is $\mathbf{x}^0_0$ for $T_{h,0}$ and $\mathbf{x}^0_1$ for $T_{h,1}$ at the same position. The red colour refers to $T_{h,0}$, the black colour refers to $T_{h,1}$. $\psi^0_0$ and $\psi^0_1$ are the test-functions at the membrane coordinate $\mathbf{x}^0_0$ and $\mathbf{x}^0_1$, respectively. Adding up both one-sided test-functions by addition of corresponding matrix lines effectively mimics that the momentum equations were tested with a single test-function of an interfacially connected grid.}
\end{figure}

At the coordinate of each membrane DOF $j$ there exist exactly two P2-test-functions $\psi_0^j, \psi_1^j$ which are non-zero, with $\operatorname{supp}(\psi_0^j)\subset\Omega_0 $ and $\operatorname{supp}(\psi_1^j)\subset\Omega_1 $.
These test-functions are each used twice, once for testing with the $v_x$-equation and once for testing with the $v_r$-equation.
Each of these test cases corresponds to a single row in the assembled Finite-Element matrix. The correct stress jump condition can be realized by adding the two rows related to $\psi_1^j$ to the two respective rows related to $\psi_0^j$, for every $j=0,\dots,n-1$. 
This adds both one-sided test-functions and effectively mimics that the momentum equations were tested with a single test-function of an interfacially connected grid. Also the boundary conditions \autoref{eq:bc_stress0} and \autoref{eq:bc_stress1} are effectively added up to recover the correct stress jump condition \autoref{eq:boundaryConditions}. As an illustration, \autoref{fig:testFunctions} shows the test-functions in the case of an 1D mesh. 

The obsolete rows related to $\psi_1^j$ can then be overwritten to enforce the continuity of the velocities at the interface
\begin{align}
\mathbf{v}_1^j - \mathbf{v}_0^j &= 0, &&\text{on }\Gamma_{h,1}.\label{eq:boundaryVelocityConstant}
\end{align}
As ${\bf v}=(v_x,v_r)$ the above describes two continuity conditions which can be assembled in the $v_x$- and $v_r$-rows tested with $\psi_1^j$.
The continuity of velocity ensures in particular, that corresponding DOFs on $\Gamma_{h,0}$ and $\Gamma_{h,1}$ share the same coordinate points for all times.

\subsection{Implementation details II - extension of surface equations to $\Omega$}
In AMDiS, it is not possible to solve 2D bulk equations and 1D surface equations all in one system. To compute the principal stretches implicitly, it is therefore necessary to extend \autoref{eq:principalStretchesImplicit1} to $\Omega$. The equations for the principal stretches then read
\begin{align}
\partial_t\lambda_1 - \lambda_1\nabla_{\Gamma}\cdot\mathbf{v}_{\Gamma,\text{ext}} = 0,\quad   \partial_t\lambda_2 - \lambda_2 \frac{(v_{\Gamma,\text{ext}})_r}{R_{\text{ext}}} = 0\,  \quad \text{in }\Omega,
\label{eq:principalStretchesInOmega}
\end{align}
where $\mathbf{v}_{\Gamma,\text{ext}}$ is the extension of the interfacial velocity to $\Omega$ constant in normal direction. $R_{\text{ext}}$ is the distance to the symmetry axis for any point in $\Omega$. The equations for the principal stretches are solved using $P1$ elements. The calculation of the normal extension is based on a Hopf-Lax algorithm described in \cite{stocker2005amdis}. The surface divergence then reads
\begin{align}
\nabla_{\Gamma}\cdot\mathbf{v}_{\Gamma,\text{ext}} = \mathbf{P}:\nabla\mathbf{v}_{\Gamma,\text{ext}},
\end{align}
with the projection onto the tangent space $\mathbf{P} = \mathbf{I} - \mathbf{n}\mathbf{n}^T$.

  \bibliographystyle{elsarticle-num} 
  \bibliography{bibliographie}




\end{document}